\documentclass[onecolumn,notitlepage,letterpaper,superscriptaddress,nofootinbib,longbibliography]{revtex4-2}

\usepackage[english]{babel}
\usepackage[T1]{fontenc}
\usepackage[utf8]{inputenc}
\usepackage{amsmath,accents}
\usepackage{amsfonts,color}
\usepackage{amssymb,float}
\usepackage{mathptmx}

\usepackage{graphicx}
\usepackage{subfigure}
\usepackage{color}

\usepackage[pdftoolbar = true, pdfstartview = FitH, pdfmenubar = true]{hyperref}
\hypersetup{
    colorlinks=true,
    linkcolor=blue,
    filecolor=magenta,      
   citecolor=blue
}

\newcommand{\lc}[1]{\accentset{\circ}{#1}}

\newcommand{\dd}{{\rm d}}

\DeclareMathOperator{\arccosh}{arccosh}

\begin{document}

\title{Coincident gauge for static spherical field configurations in symmetric teleparallel gravity}

\author{Sebastián Bahamonde}
\email{bahamonde.s.aa@m.titech.ac.jp}
\affiliation{Department of Physics, Tokyo Institute of Technology,
1-12-1 Ookayama, Meguro-ku, Tokyo 152-8551, Japan}
\author{Laur Järv} 
\email{laur.jarv@ut.ee}
\affiliation{Institute of Physics, University of Tartu, W.\ Ostwaldi 1, 50411 Tartu, Estonia}



\begin{abstract}In symmetric teleparallel gravities, where the independent connection is characterized by nonmetricity while curvature and torsion are zero, it is possible to find a coordinate system whereby the connection vanishes globally and covariant derivatives reduce to partial derivatives -- the coincident gauge. In this paper we derive general transformation rules into the coincident gauge for spacetime configurations where the both the metric and connection are static and spherically symmetric, and write out the respective form of the coincident gauge metrics. Taking different options in fixing the freedom in the connection allowed by the symmetry and the field equations, the Schwarzschild metric in the coincident gauge can take for instance the Cartesian, Kerr-Schild, and diagonal (isotropic-like) forms, while the BBMB black hole metric in symmetric teleparallel scalar-tensor theory a certain diagonal form fits the coincident gauge requirements but the Cartesian and Kerr-Schild forms do not. Different connections imply different value for the boundary term which could in principle be physically relevant, but simple arguments about the coincident gauge do not seem to be sufficient to fix the connection uniquely. As a byproduct of the investigation we also point out that only a particular subset of static spherically symmetric connections has vanishing nonmetricity in the Minkowski limit.
\end{abstract}

\maketitle

\section{Introduction}

It is an elementary textbook maxim that when going from the special relativity formulae in Cartesian coordinates of Minkowski space into the general relativity (GR) expressions in the language of Riemannian geometry, the partial derivatives shall be promoted to covariant derivatives. The procedure introduces Levi-Civita affine connection given as Christoffel symbols, which are mathematically defined as the derivatives of the basis vectors, and physically represent the inertial and gravitational forces experienced by a particle (indistinguishable by the virtue of the equivalence principle). On a Riemannian manifold it is always possible to make a transformation into the normal coordinates, where the affine connection vanishes at a single point. At this point the covariant derivatives reduce to the partial derivatives and the geodesic motion of a particle exhibits no acceleration. Despite the economy achieved by the vanishing connection, practical computations in the normal coordinates are not necessarily simpler, since the metric could obtain a rather cumbersome form instead. These properties are inherited by the extended theories like $f(\lc{R})$ and scalar-tensor gravity as well, as ingrained in the Riemannian geometry.

The state of affairs gets more interesting, however, in the non-Riemannian framework where the affine connection is taken to be independent of the metric, and thus besides curvature can be characterized by torsion and nonmetricity as well. Assuming flatness, i.e., that the curvature vanishes, i.e., the geometry is teleparallel (as the orientation of parallel transported vectors does not change), poses no obstacle for a meaningful theory of gravity. In the contrary, since the Levi-Civita curvature scalar $\lc{R}$ of the Einstein-Hilbert action can be rewritten in terms of the torsion scalar $T$ and a boundary term $B_T$, or the nonmetricity scalar $Q$ and a boundary term $B_Q$, the dynamical content of GR can be transplanted into the setting of metric teleparallelism (with zero curvature and zero nonmetricity) or symmetric teleparallelism (with zero curvature and zero torsion). The respective theories are known as the teleparallel equivalent of general relativity (TEGR) \cite{Aldrovandi:2013wha,Krssak:2018ywd} and symmetric teleparallel equivalent of general relativity (STEGR) \cite{Nester:1998mp,BeltranJimenez:2017tkd}, which together with GR form a ``geometrical trinity of gravity'' \cite{BeltranJimenez:2018vdo,BeltranJimenez:2019tjy}, or rather inhabit a full circle of equivalent theories \cite{BeltranJimenez:2019odq}. Analogously to the Riemannian case, the teleparallel theories can be extended by considering a gravitational action depending on a general function of the ruling geometric quantity like $f(T)$ \cite{Bengochea:2008gz,Linder:2010py,Golovnev:2017dox,Hohmann:2017duq} or $f(Q)$~\cite{BeltranJimenez:2017tkd,Koivisto:2019ggr} gravity, or by introducing a nonminimal coupling between a scalar field and this quantity, thus yielding the scalar-tensor versions of teleparallel gravity \cite{Geng:2011aj,Hohmann:2018rwf} and symmetric teleparallel gravity \cite{Jarv:2018bgs}. More elaborate extensions have been considered as well (see Ref.\ \cite{Bahamonde:2021gfp} for a comprehensive review). In general the field equations of these extended theories are different from their counterparts in the extensions of the original formulation of GR, thus providing an entirely new landscape for study and investigation. In this paper, we will be focused on studying the symmetric teleparallel framework where only non-metricity is present.

The extra non-Riemannian freedom in connection leads to a theorem that in the case of zero curvature and zero nonmetricity, the connection can be made to vanish globally in a set of anholonomic (local Lorentz) frames, while in the case of zero curvature and zero torsion in a set of holonomic frames \cite{Eisenhart:1927}. In other words, metric teleparallelism allows a global tetrad field whereby the associated spin connection vanishes, and symmetric teleparallelism allows a coordinate system whereby the affine connection vanishes everywhere on a manifold, not just at a point. In the literature the former is often called a ``Weitzenböck gauge'', while following Ref.\ \cite{BeltranJimenez:2017tkd} the latter is known as a ``coincident gauge'', since the vanishing of the connection makes the covariant and partial derivatives to coincide.

In the metric (torsional) version of teleparallelism, the Weitzenböck gauge tetrad ansatz for a static spherically symmetric spacetime was first found by carefully solving the field equations in extended $f(T)$ gravity \cite{Ferraro:2011ks,Tamanini:2012hg}, and only later it was realized that in this form the tetrad, and consequently the independent connection obeys the same symmetry as the metric \cite{Hohmann:2019nat}. Similarly, the initially found Weitzenböck gauge solutions of rotating tetrads \cite{Bejarano:2014bca,Jarv:2019ctf} were later seen to belong to the two axially symmetric branches of connections \cite{Hohmann:2019nat}, out of which only one branch has a proper spherically symmetric limit \cite{Bahamonde:2020snl}. Thus the assumption of the connection being endowed with the same symmetry as the metric seems a useful premise in approaching the field configurations.

The torsional version of teleparallel gravity has been studied in different physical situations but the family of symmetric (nonmetricity) teleparallel gravities is less explored, and only a few examples of the coincident gauge have been explicitly written down. It was noticed from early on that besides the obvious Minkowski case the spatially flat Friedmann-Lema\^itre-Robertson-Walker metric in Cartesian coordinates is compatible with vanishing connection \cite{BeltranJimenez:2017tkd}. Actually, asking the connection to obey the cosmological symmetry of isotropy and homogeneity yields three different sets of connections for the spatially flat case and one set for the spatially hyperbolic and spherical geometries \cite{Hohmann:2021ast,DAmbrosio:2021pnd} (all showing different phenomenology \cite{Dimakis:2022rkd}), while only for one of the spatially flat sets going to the coincident gauge reduces to the reparametrization of the time coordinate, i.e.\ is congruent with the Cartesian coordinates \cite{Hohmann:2021ast}. These results also hold in the reverse order, i.e.\ first assuming a coincident gauge and then imposing the symmetry of the connection \cite{BeltranJimenez:2022azb}. For static spherically symmetric spacetime the usual spherical coordinates are not consistent with the coincident gauge \cite{Zhao:2021zab,Lin:2021uqa}, while the Cartesian \cite{Zhao:2021zab,Lin:2021uqa} and Kerr-Schild \cite{BeltranJimenez:2019bnx,Gomes:2022vrc} coordinates can be compatible with the vanishing connection.

In this paper we aim to further the understanding of the coincident gauge and derive the corresponding general transformation rules as well as the respective metrics for static spherically symmetric field configurations. We assume that both the metric and connection obey the same symmetry, and start from the two sets of curvature and torsion-free connections worked out recently \cite{DAmbrosio:2021zpm}. In Sec.~\ref{sec:action} we give a brief introduction to symmetric teleparallel gravity and also provide the action and field equations for a class of
scalar-tensor theory based on a non-minimal coupling between a scalar field and the nonmetricity scalar. We elaborate the implications
of the coincident gauge as well. Sec.~\ref{sec:spheri} is devoted to explain static spherical symmetry for any generic theory based in symmetric teleparallel gravity by assuming that the connection is also spherically symmetric. The most central part of the paper is presented in Sec.~\ref{sec:coincidentG} where we solve the coincident gauge transformations for the different sets and give the general form of the respective coincident gauge metrics. In Sec.\ \ref{sec:Mink} we point out that only a specific branch of the spherical static connections has vanishing nonmetricity in the Minkowski limit, and focus on this branch in the subsequent explicit examples. To illustrate the results we consider the Schwarzschild solution in Sec.\ \ref{sec:Schwarzschild} and show that Cartesian, Kerr-Schild, as well as isotropic-like coordinates can fit coincident gauge, but each of these choices fixes the original freedom in the connection differently. As the second example we take the so-called BBMB solution in symmetric teleparallel scalar-tensor theory, and find that neither Cartesian nor Kerr-Schild coordinates can provide the coincident gauge, but certain isotropic coordinates do.
Finally, we give a brief conclusion about our main findings in Sec.~\ref{sec:conclusions}. 

Our convention is the following: we use units where $c=1$ and the metric signature $(-,+,+,+)$. We denote Levi-Civita quantities (Riemannian) with a circle on top $\circ$ and quantities without any symbol would be refer to be symmetric teleparallel gravity.

\section{Symmetric teleparallel gravities}\label{sec:action}

Symmetric teleparallel gravity assumes a geometric setup where the connection is characterized by identically vanishing curvature and torsion, while only nonmetricity is left to carry nontrivial information. The action can be constructed from the nonmetricity scalar which is equivalent to the Levi-Civita Ricci scalar up to a boundary term, thus providing a link to GR. It is a special property of these geometries that there exists a system of coordinates where the connection can be made to vanish globally, a so-called coincident gauge.

\subsection{Geometric preliminaries}

A generic connection $\tilde{\Gamma}{}^{\lambda}_{\phantom{\alpha}\mu\nu}$ with 64 independent components can be decomposed into three parts, \cite{Hehl:1976kj,Hehl:1994ue}
\begin{equation}
\label{Connection decomposition}
\tilde{\Gamma}{}^{\lambda}_{\phantom{\alpha}\mu\nu} = 
\lc{\Gamma}^{\lambda}{}_{\mu\nu} +
K^{\lambda}_{\phantom{\alpha}\mu\nu}+
 L^{\lambda}_{\phantom{\alpha}\mu\nu} \,,
\end{equation}
namely the Levi-Civita connection of the metric $g_{\mu\nu}$,
\begin{equation}
\label{LeviCivita}
 \lc{\Gamma}^{\lambda}{}_{\mu \nu} \equiv \frac{1}{2} g^{\lambda \beta} \left( \partial_{\mu} g_{\beta\nu} + \partial_{\nu} g_{\beta\mu} - \partial_{\beta} g_{\mu\nu} \right) \,,
\end{equation}
contortion tensor
\begin{equation}
\label{Contortion}
 K^{\lambda}{}_{\mu\nu} \equiv \frac{1}{2} g^{\lambda \beta} \left( -T_{\mu\beta\nu}-T_{\nu\beta\mu} +T_{\beta\mu\nu} \right) =-\,K_{\mu}{}^{\lambda}{}_{\nu}\, ,
\end{equation}
and disformation tensor
\begin{equation}
\label{Disformation}
L^{\lambda}{}_{\mu\nu} \equiv \frac{1}{2} g^{\lambda \beta} \left( -Q_{\mu \beta\nu}-Q_{\nu \beta\mu}+Q_{\beta \mu \nu} \right) = L^{\lambda}{}_{\nu\mu}  \,.
\end{equation}
Here contortion is built from torsion tensors
\begin{equation}
\label{TorsionTensor}
 T^{\lambda}{}_{\mu\nu}\equiv \tilde{\Gamma}{}^{\lambda}{}_{\mu\nu}-\tilde{\Gamma}{}^{\lambda}{}_{\nu\mu}\,,
\end{equation}
while disformation is constructed from nonmetricity tensors
\begin{equation}
\label{NonMetricityTensor}
Q_{\rho \mu \nu} \equiv \nabla_{\rho} g_{\mu\nu} = \partial_\rho g_{\mu\nu} - \tilde{\Gamma}{}^\beta{}_{\mu \rho} g_{\beta \nu} -  \tilde{\Gamma}{}^\beta{}_{\nu \rho } g_{\mu \beta}  \,.
\end{equation}
The latter two along with curvature tensor
\begin{eqnarray}
\label{CurvatureTensor}
    R^{\sigma}\,_{\rho\mu\nu} &\equiv& \partial_{\mu}\tilde{\Gamma}{}^{\sigma}\,_{\nu\rho}-\partial_{\nu}\tilde{\Gamma}{}^{\sigma}\,_{\mu\rho}+\tilde{\Gamma}{}^{\sigma}\,_{\mu\lambda}\tilde{\Gamma}{}^{\lambda}\,_{\nu\rho}-\tilde{\Gamma}{}^{\sigma}\,_{\mu\lambda}\tilde{\Gamma}{}^{\lambda}\,_{\nu\rho}\,
\end{eqnarray}
are the three key properties that characterize a connection. When curvature is zero, the orientation of vectors does not change under parallel transport along a curve. When torsion is zero, the connection is symmetric in the lower indices. Hence the imposition of vanishing curvature and torsion warrants the name ``symmetric teleparallel'', and we denote it by ${\Gamma}{}^{\lambda}_{\phantom{\alpha}\mu\nu}$.

It is an interesting property of the connection ${\Gamma}{}^{\lambda}_{\phantom{\alpha}\mu\nu}$ that the scalar curvature of the Levi-Civita part of the connection, $\lc{R}$, can be expressed as the sum of a nonmetricity scalar and a Levi-Civita divergence acting over the two independent traces of the nonmetricity tensor, 
\begin{eqnarray}\label{R and Q}
\lc{R}=Q+\lc{\nabla}_{\mu}(\hat{Q}^\mu-Q^\mu) \,,
\end{eqnarray}
where the nonmetricity scalar and traces are defined as
\begin{align}\label{Qscalar}
    Q &\equiv -\,\frac{1}{4}\,Q_{\lambda\mu\nu}Q^{\lambda\mu\nu}+\frac{1}{2}\,Q_{\lambda\mu\nu}Q^{\mu\nu\lambda}+\frac{1}{4}\,Q_{\mu}Q^{\mu}-\frac{1}{2}\,Q_{\mu}\hat{Q}^{\mu}\,, \\
    Q_{\mu} &\equiv Q_{\mu\nu}\,^{\nu}\,, \qquad \qquad
    \hat{Q}_{\mu} \equiv Q_{\nu\mu}\,^{\nu}\,.
\end{align}
It is also significant that the total divergence part in \eqref{R and Q}, 
\begin{align}
\label{eq: B_Q}
    B_Q &=\lc{\nabla}_{\mu}(\hat{Q}^\mu-Q^\mu) 
\end{align} 
becomes a boundary term under a spacetime integral.

\subsection{Action and field equations}

As the Einstein-Hilbert action of GR is given by the Levi-Civita curvature scalar $\lc{R}$, we can rewrite that action using the nonmetricity scalar $Q$ instead, and expect to keep the same dynamical content since the boundary term does not affect the field equations. This is the idea behind symmetric teleparallel equivalent of general relatvity. Various extensions of the symmetric teleparallel theory like substituting $Q$ in the action by $f(Q)$ or introducing a nonminimal coupling between $Q$ and a scalar field $\Phi$, however, lead to theories that are different from their counterparts $f(\lc{R})$ and scalar-tensor gravity originally formulated in the Riemannian geometry.

A relatively simple, but still versatile extended symmetric teleparallel action can be written as \cite{Jarv:2018bgs}
\begin{equation}
\label{Action}
S = \frac{1}{2\kappa^2} \int\mathrm{d}^4x \sqrt{-g} \left( \mathcal{A}(\Phi) Q -\mathcal B(\Phi) g^{\alpha\beta}\partial_\alpha\Phi \partial_\beta\Phi 
-2{\mathcal V}(\Phi)\right) + S_{\mathrm{m}}\,,
\end{equation}
where $\kappa^2=8\pi G$ and $S_{\mathrm{m}}$ denotes the action of matter (which we assume to be the same as in GR, i.e.\ depending on the metric alone). 
Like in the usual Riemannian scalar-tensor theory the nonminimal coupling function $\mathcal{A}$ sets the strength of the effective gravitational constant, $\mathcal{B}$ is the kinetic coupling function, and $\mathcal{V}$ is the scalar potential. If the nonminimal coupling function is fixed to unity, $\mathcal{A}(\Phi)\equiv 1$, and the kinetic and potential terms of the scalar field vanish, $\mathcal{B}(\Phi)\equiv\mathcal{V}(\Phi)\equiv 0$, the theory is reduced to a STEGR. If the nonminimal coupling function is unity, but and the kinetic term of the scalar field is nontrivial, then we have a theory that is equivalent to a minimally coupled scalar field in GR. In fact, also the better known $f(Q)$ gravity can be viewed as a special subcase of the scalar-tensor action \eqref{Action}, as it can be obtained by setting~\cite{Jarv:2018bgs}
\begin{align}\label{transftofQ}
 \mathcal{A}=f_Q\,,\qquad
 \mathcal{B}=0\,,\qquad
 \mathcal{V}=\tfrac{1}{2} \left(Q f_Q-f\right)\,,\qquad \Phi=Q\,,
\end{align}
where $f_Q=df/dQ$.

With the help of introducing the so-called superpotential (or conjugate) tensor 
\begin{eqnarray}
 P^\alpha{}_{\mu\nu}=-\,\frac{1}{4}Q^{\alpha}{}_{\mu\nu}+\frac{1}{2}Q_{(\mu}{}^\alpha{}_{\nu)}+\frac{1}{4}g_{\mu\nu}Q^\alpha-\frac{1}{4}(g_{\mu\nu}\hat{Q}^\alpha+\delta^\alpha{}_{(\mu}Q_{\nu)})\,
\end{eqnarray}
as well as some geometric indentities the field equations arising from the variation of the action \eqref{Action} with respect to the metric, symmetric teleparallel connection, and scalar field are \cite{Jarv:2018bgs,Bahamonde:2022esv}
\begin{subequations}
\label{eq: scalar-tensor field equations}
\begin{eqnarray}
\mathcal{A}(\Phi)\lc{G}_{\mu\nu}
+2\frac{\dd\mathcal{A}}{\dd\Phi} P^\lambda{}_{\mu\nu} \partial_\lambda \Phi + \frac{1}{2} g_{\mu\nu} \left( \mathcal{B}(\Phi) g^{\alpha\beta}\partial_\alpha\Phi \partial_{\beta} \Phi + 2 {\mathcal V}(\Phi)  \right) - \mathcal{B}(\Phi) \partial_{\mu} \Phi \partial_\nu \Phi &=&  \kappa^2\mathcal{T}_{\mu\nu} \,,
 \label{MetricFieldEqF}
\\
\label{connectionEq}
\left( \frac{1}{2}Q_\beta + \nabla_\beta \right) \left[ \partial_\alpha \mathcal{A} \left( \frac{1}{2}Q_\mu g^{\alpha\beta}-\frac{1}{2}\delta^\alpha_\mu Q^\beta-Q_{\mu}\,^{\alpha\beta}+\delta^\alpha_\mu Q_{\gamma}\,^{\gamma\beta}\right)\right] &=&0\,,
\\
\label{ScalarFieldEq}
2\mathcal B \lc{\square}\Phi 
+\frac{\dd\mathcal{B}}{\dd\Phi}g^{\alpha\beta}\partial_{\alpha}\Phi\partial_{\beta}\Phi + \frac{\dd\mathcal{A}}{\dd\Phi} Q -
2\frac{\dd\mathcal{V}}{\dd\Phi} &=&0 \,.
\end{eqnarray}
\end{subequations}
Here $\lc{G}_{\mu\nu}$ is the Einstein tensor and $\lc{\square}$ the d'Alembert operator computed of the  Levi-Civita part of the connection, while $\mathcal{T}_{\mu\nu}$ is the usual matter energy-momentum tensor. When the scalar field is globally constant, the Eq.\ \eqref{MetricFieldEqF} reduces to the Einstein's equation in GR with the value of the potential playing the role of the cosmological constant, while \eqref{connectionEq} and \eqref{ScalarFieldEq} immediately vanish. Therefore, the solutions of GR are trivially also the solutions of these scalar-tensor theories with the scalar field being constant. 

One can also notice that if the nonminimal coupling function $\mathcal{A}(\Phi)$ is constant, the connection field equation \eqref{connectionEq} is identically satisfied and all contributions of the nonmetricity drop out in the remaining equations as well. Thus in the case of STEGR, the non-Riemannian part of the symmetric teleparallel connection is left completely undetermined by the field equations. This feature can be traced back to the identity \eqref{R and Q}, which implies that the difference between GR and STEGR is at most in the boundary term only. The boundary term does not concern the field equations, and thus the STEGR field equations can not contain anything extra to the GR field equations. As the GR field equations know only about the Riemannian part of the connection, the non-Riemannian part of the symmetric teleparallel connection can not emerge in the STEGR field equations either. 
However, it might still be possible to entertain some heuristic arguments that could be used in fixing the non-Riemannian part of the connection nevertheless, relevant in modified theories of gravity or in physical situations when the value of the action plays a role (such as in black hole entropy~\cite{Gomes:2022vrc}).

\subsection{Coincident gauge}

The action \eqref{Action} and the ensuing field equations are manifestly covariant, in the sense that under a general coordinate transformation from $\xi^\lambda$ into $x^\lambda$ the metric and connection transform as usual
\begin{subequations}
\label{transf}
\begin{eqnarray}
\label{eq: metric transformation}
{g}_{\mu\nu}(x^\lambda) &=& \frac{\partial \xi^\alpha}{\partial x^\mu}\frac{\partial \xi^\beta}{\partial x^\nu}g_{\alpha\beta} (\xi^\lambda) \,, \\
\label{eq: Gamma transformation}
\Gamma^\rho{}_{\mu\nu}(x^\lambda) &=& \frac{\partial x^\rho}{\partial \xi^\gamma} \frac{\partial \xi^\alpha}{\partial x^\mu}\frac{\partial \xi^\beta}{\partial x^\nu}\Gamma^\gamma{}_{\alpha\beta} (\xi^\lambda) + \frac{\partial^2 \xi^\alpha}{\partial x^\mu \partial x^\nu } \frac{\partial x^\rho}{\partial \xi^\alpha} \,.
\end{eqnarray} 
\end{subequations}
In particular, the quantities $\lc{R}$, $Q$, and $B_Q$ transform as scalars and retain their relationship \eqref{R and Q}.

A very interesting mathematical fact in the case of symmetric teleparallel geometry is that there exists a coordinate system where the connection vanishes globally on the manifold \cite{Eisenhart:1927}. The authors of Ref.\ \cite{BeltranJimenez:2017tkd} have dubbed this system of coordinates a ``coincident gauge'', since the vanishing of the connection makes the covariant and partial derivatives to coincide. Also then the nonmetricity tensor \eqref{NonMetricityTensor} reduces to the simple form of partial derivatives of the metric only. When the coordinate system corresponding to the coincident gauge is known, the transformation \eqref{eq: Gamma transformation} of the connection into any other coordinate system $x^\alpha$ reduces to
\begin{equation}
\label{eq: coincident gauge}
    \Gamma^\rho{}_{\mu \nu}(x^\lambda) = \left(\frac{\partial \xi^\alpha}{\partial x^\rho}\right)^{-1} \partial_\mu \left(\frac{\partial \xi^\alpha}{\partial x^\nu}\right) \,,
\end{equation}
where $\xi^\lambda$ are the coordinates of the coincident gauge with $\Gamma^\gamma{}_{\alpha\beta}(\xi^\lambda)=0$. Similarly, by an inverse transformation of \eqref{transf} it is possible from any other coordinate system to transform back into the coincident gauge (provided the determinant of the Jacobian is not zero).

The combined property that the non-Riemannian part of the connection is not determined by the STEGR field equations and that there exists a system of coordinates where the teleparallel connection vanishes may tempt one to entertain a noncovariant approach to STEGR by just imposing vanishing teleparallel connection in association with any metric. In this approach under the coordinate transformations one would only transform the metric \eqref{eq: metric transformation} while instead of \eqref{eq: Gamma transformation} the connection is kept always zero. (In analogy with the metric teleparallel case \cite{Blixt:2022rpl} one may speak of a ``pure metric'' approach to symmetric teleparallel gravity and call this a coordinate transformation of the 2nd kind). Transforming only the metric is equivalent to making a combined transformation \eqref{transf} and on top of that only the connection transformation (inverse of \eqref{eq: coincident gauge}) to make the connection to vanish. While such transformations are consistent with the STEGR field equations, it is rather questionable whether they provide a true physical symmetry of the theory which relates physically equivalent configurations (as the usual coordinate transformations \eqref{transf} do). Under the pure metric transformation \eqref{eq: metric transformation} the nonmetricity tensor would transform noncovariantly (since $\partial_\rho g_{\mu\nu}$ is not a tensor), and the nonmetricity scalar would not remain at the same value but pick up an extra contribution in the form of an additional boundary term. Thus the relation \eqref{R and Q} would be altered in the sense that $\lc{R}$ and $Q$ would be related to each other by a different boundary term before and after the transformation. The field equations would still assume the same covariant form, but any quantity that relies on the boundary term (like the black hole entropy) would report altered physics.

In our view Eq.\ \eqref{eq: coincident gauge} is rather a generating rule of a set of all possible symmetric teleparallel connections, where the connection corresponding to a particular physical configuration is one member. Indeed, for a given metric $g_{\mu\nu}(x^\lambda)$ picking a set of all possible four functions $\xi^\alpha(x^\lambda)$ generates a set of all connections that have the property of being curvature and torsion-free. Connection components which are solutions of symmetric teleparallel field equations necessarily belong to this set. In STEGR this whole set is consistent with the field equations since the STEGR field equations leave the non-Riemannian part of the connection undetermined, while for the extended theories like scalar-tensor the field equations will in general fix the connection or at least constrain it to a particular subset of the curvature and torsion-free set. An obvious question remains how to fix the connection for a physical configuration when the field equations are not able to fully settle it? In the next section we are are going to see how the demand that the connection obeys the same symmetry as the metric which constrains the set a bit further. This provides the stage for a transformation into the coincident gauge, where different heuristic considerations can be tried out.

\section{Static spherically symmetric metric and connection}\label{sec:spheri}

In the spherical coordinates $x^\mu=(t,r,\theta,\phi)$ static and spherically symmetric spacetimes are characterized by the Killing vectors $Z_\zeta$
\begin{subequations}
\label{eq: Killing spherical set 1}
\begin{align}
K^\mu &= \begin{pmatrix} 1 & 0 & 0 & 0 \end{pmatrix}  \,, \\
R^\mu &= \begin{pmatrix} 0 & 0 & 0 & 1 \end{pmatrix} \,, \\
S^\mu &= \begin{pmatrix} 0 & 0 & \cos\phi & -\sin\phi \cot \theta \end{pmatrix} \,, \\
T^\mu &= \begin{pmatrix} 0 & 0 & -\sin\phi & -\cos{\phi}\cot{\theta} \end{pmatrix} \,.
\end{align}
\end{subequations}
Imposing the symmetry means that the Lie derivatives of the metric and affine connection along these vectors vanish,
\begin{align}\label{LieD_mag}
\mathcal{L}_{Z_\zeta}g_{\mu\nu}=0\,,\qquad
\mathcal{L}_{Z_\zeta}\tilde{\Gamma}^{\lambda}\,_{\mu\nu}=0 \,.
\end{align}
Note that we have extra imposed that the general affine connection respects the same symmetries as the metric tensor.
For the metric this gives the well known result
\begin{equation}
\label{eq: spherical metric}
    ds^2=-g_{tt}(r) \, dt^2+g_{rr}(r) \, dr^2+ r^2 d\theta^2 + r^2 \sin^2\theta \, d\phi^2\,,
\end{equation}
where $g_{tt}$, $g_{rr}$ are two arbitrary functions of the radial coordinate and must be determined by the field equations of the theory at hand. General affine connection that obeys this symmetry, however, is parameterized by twenty functions $\mathcal{C}_n(r)$ in the form \cite{Hohmann:2019fvf}
\begin{align}
{\tilde{\Gamma}}{}^{\rho}{}_{\mu\nu} &=\left[\begin{matrix}
   \left[\begin{matrix}\mathcal{C}_1 & \mathcal{C}_2 & 0 & 0\\\mathcal{C}_3 & \mathcal{C}_4 & 0 & 0\\0 & 0 & \mathcal{C}_9 & -\mathcal{C}_{19} \sin \theta\\0 & 0 & \mathcal{C}_{19} \sin \theta & \mathcal{C}_{9} \sin^2 \theta \end{matrix}\right] & 
   \left[\begin{matrix} \mathcal{C}_5 & \mathcal{C}_6 & 0 & 0\\\mathcal{C}_7 & \mathcal{C}_8 & 0 & 0\\0 & 0 & \mathcal{C}_{10} & -\mathcal{C}_{20} \sin \theta\\0 & 0 & \mathcal{C}_{20} \sin \theta & \mathcal{C}_{10} \sin^2 \theta \end{matrix}\right] & 
    \end{matrix}\right. \nonumber \\
&  \qquad \left.\begin{matrix} & 
   \left[\begin{matrix}0 & 0 & \mathcal{C}_{11} & -\mathcal{C}_{15}\sin\theta\\0 & 0 & \mathcal{C}_{12} & -\mathcal{C}_{16}\sin\theta\\ \mathcal{C}_{13} & \mathcal{C}_{14} & 0 & 0\\ -\mathcal{C}_{17}\sin\theta & -\mathcal{C}_{18}\sin\theta & 0 & - \sin{\theta} \cos{\theta}\end{matrix}\right] &
   \left[\begin{matrix}0 & 0 & \vspace{0.1cm}\frac{\mathcal{C}_{15}}{\sin\theta} & \mathcal{C}_{11}\\\vspace{0.1cm} 0 & 0 & \frac{\mathcal{C}_{16}}{\sin\theta} & \mathcal{C}_{12}\\ \vspace{0.1cm}\frac{\mathcal{C}_{17}}{\sin\theta} & \frac{\mathcal{C}_{18}}{\sin\theta} & 0 & \cot{\theta}\\\mathcal{C}_{13} & \mathcal{C}_{14} & \cot{\theta} & 0\end{matrix}\right]\end{matrix}\right]
   \label{eq: Gamma set10}\,.
\end{align}
Here the notation should be understood as ${\tilde{\Gamma}}{}^{\rho}{}_{\mu\nu} = \left[ \begin{matrix} {\tilde{\Gamma}}{}^{0}{}_{\mu\nu} & {\tilde{\Gamma}}{}^{1}{}_{\mu\nu} & {\tilde{\Gamma}}{}^{2}{}_{\mu\nu} & {\tilde{\Gamma}}{}^{3}{}_{\mu\nu} \end{matrix} \right]$ where in the four blocks the $\mu$ and $\nu$ indices label the rows and columns, respectively. Demading that the affine connection is curvature and torsion-free reduces the number of free functions considerably. It turns out there are two solutions, both having two independent functional freedoms left \cite{DAmbrosio:2021zpm}. 

\subsection{Connection set 1}

The set 1 of static and spherically symmetric curvature and torsion-free connections is given by~\cite{DAmbrosio:2021zpm}
\begin{align}
\Gamma^{\rho}{}_{\mu\nu} &=
   \left[\begin{matrix}\left[\begin{matrix}c & \Gamma^{\phi}{}_{r \phi} & 0 & 0\\\Gamma^{\phi}{}_{r \phi} & \Gamma^{t}{}_{rr} & 0 & 0\\0 & 0 & - \frac{1}{c} & 0\\0 & 0 & 0 & - \frac{\sin^{2}{\theta}}{c}\end{matrix}\right] & \left[\begin{matrix}0 & 0 & 0 & 0\\0 & \Gamma^{r}{}_{rr} & 0 & 0\\0 & 0 & 0 & 0\\0 & 0 & 0 & 0\end{matrix}\right] & \left[\begin{matrix}0 & 0 & c & 0\\0 & 0 & \Gamma^{\phi}{}_{r \phi} & 0\\c & \Gamma^{\phi}{}_{r \phi} & 0 & 0\\0 & 0 & 0 & - \sin{\theta} \cos{\theta}\end{matrix}\right] \end{matrix} 
   \begin{matrix} \left[\begin{matrix}0 & 0 & 0 & c\\0 & 0 & 0 & \Gamma^{\phi}{}_{r \phi}\\0 & 0 & 0 & \cot{\theta}\\c & \Gamma^{\phi}{}_{r \phi} & \cot{\theta} & 0\end{matrix}\right]\end{matrix}\right]
   \label{eq: Gamma set1}
\end{align}
where the functions $\Gamma^{\phi}{}_{r \phi}$, $\Gamma^{t}{}_{rr}$, $\Gamma^{r}{}_{rr}$ depend on $r$, while $c$ is a nonzero constant. The functions have to obey the relation
\begin{align}
    \frac{d}{d r} \Gamma^{\phi}{}_{r \phi} &= c \Gamma^{t}{}_{rr} - \Gamma^{\phi}{}_{r \phi} ( \Gamma^{\phi}{}_{r \phi} + \Gamma^{r}{}_{rr}) \,.
    \label{eq: Gamma relations set1}
\end{align}
to satisfy vanishing curvature and torsion, hence overall there are two functional degrees of freedom.

\subsection{Connection set 2}

The set 2 of static and spherically symmetric curvature and torsion-free connections is given by~\cite{DAmbrosio:2021zpm}
\begin{align}
\Gamma^{\rho}{}_{\mu\nu} =&
\left[\left[\begin{matrix}- c \left(2 c - k\right) \Gamma^{t}{}_{\theta \theta}  - c + k & \frac{\left(2 c - k\right) \left(c \Gamma^{t}{}_{\theta \theta}  + 1\right) \Gamma^{t}{}_{\theta \theta} }{\Gamma^{r}{}_{\theta \theta} } & 0 & 0\\\frac{\left(2 c - k\right) \left(c \Gamma^{t}{}_{\theta \theta}  + 1\right) \Gamma^{t}{}_{\theta \theta} }{\Gamma^{r}{}_{\theta \theta} } & \Gamma^{t}{}_{rr}  & 0 & 0\\0 & 0 & \Gamma^{t}{}_{\theta \theta}  & 0\\0 & 0 & 0 & \Gamma^{t}{}_{\theta \theta}  \sin^{2}\theta\end{matrix}\right] 
\right. \nonumber \\ & \qquad
\left[\begin{matrix}- c \left(2 c - k\right) \Gamma^{r}{}_{\theta \theta}  & c \left(2 c - k\right) \Gamma^{t}{}_{\theta \theta}  + c & 0 & 0\\c \left(2 c - k\right) \Gamma^{t}{}_{\theta \theta}  + c & \Gamma^{r}{}_{rr}  & 0 & 0\\0 & 0 & \Gamma^{r}{}_{\theta \theta}  & 0\\0 & 0 & 0 & \Gamma^{r}{}_{\theta \theta}  \sin^{2}\theta\end{matrix}\right] \nonumber \\ & \qquad \left.
\left[\begin{matrix}0 & 0 & c & 0\\0 & 0 & \frac{- c \Gamma^{t}{}_{\theta \theta}  - 1}{\Gamma^{r}{}_{\theta \theta} } & 0\\c & \frac{- c \Gamma^{t}{}_{\theta \theta}  - 1}{\Gamma^{r}{}_{\theta \theta} } & 0 & 0\\0 & 0 & 0 & - \sin\theta \cos\theta\end{matrix}\right] \quad 
\left[\begin{matrix}0 & 0 & 0 & c\\0 & 0 & 0 & \frac{- c \Gamma^{t}{}_{\theta \theta}  - 1}{\Gamma^{r}{}_{\theta \theta} }\\0 & 0 & 0 & \cot\theta\\c & \frac{- c \Gamma^{t}{}_{\theta \theta}  - 1}{\Gamma^{r}{}_{\theta \theta} } & \cot\theta & 0\end{matrix}\right] \right]
\label{eq: Gamma set2}
\end{align}
where the functions $\Gamma^{t}{}_{\theta \theta}$, $\Gamma^{t}{}_{rr}$, $\Gamma^{r}{}_{\theta \theta}$, $\Gamma^{r}{}_{rr}$ also depend on $r$, while $c$ and $k$ are some constant parameters. The functions have to obey the relations
\begin{subequations}
\label{eq: Gamma relations set2}
\begin{align}
    \frac{d}{d r} \Gamma^{t}{}_{\theta \theta}  &= - \frac{\left(\left(c \left(2 c - k\right) \Gamma^{t}{}_{\theta \theta}  + 3 c - k\right) \Gamma^{t}{}_{\theta \theta}  + 1\right) \Gamma^{t}{}_{\theta \theta} }{\Gamma^{r}{}_{\theta \theta} } - \Gamma^{r}{}_{\theta \theta}  \Gamma^{t}{}_{rr} \,, \\
    \frac{d}{d r} \Gamma^{r}{}_{\theta \theta}  &= - c \left(\left(2 c - k\right) \Gamma^{t}{}_{\theta \theta}  + 2\right) \Gamma^{t}{}_{\theta \theta}  - \Gamma^{r}{}_{\theta \theta}  \Gamma^{r}{}_{rr}  - 1 
\end{align}
\end{subequations}
for vanishing curvature and torsion, hence again there are two functional degrees of freedom remaining in the game.

\section{Coincident gauge transformations}\label{sec:coincidentG}

If the coordinates of the coincident gauge $\xi^\lambda$ are known, it is possible to make a transformation \eqref{transf} into any other coordinates $x^\lambda$, while the connection transforms as \eqref{eq: coincident gauge}. However, if we do not know the coincident gauge, we may rewrite Eq.\ \eqref{eq: coincident gauge} as
\begin{equation}
\label{eq: coincident gauge rewritten}
    \left(\frac{\partial \xi^\rho}{\partial x^\alpha}\right) \Gamma^\alpha{}_{\mu \nu}(x^\lambda) = \partial_\mu \left(\frac{\partial \xi^\rho}{\partial x^\nu}\right) \,,
\end{equation}
and use the known affine connection components in the other coordinates to solve the partial differential equations for the unknown functions $\xi^\rho(x^\lambda)$ which give the coincident gauge coordinates in terms of the known coordinates. As the requirement of static and spherical symmetry yielded two sets of connections we shall treat both of them separately. Unlinke the investigation in the cosmolgical case \cite{Hohmann:2021ast} we do not impose any particular ansatz to solve these equations. Note that to invert \eqref{eq: coincident gauge} we assumed that the transformation between the coordinate systems have the property that the determinant of the Jacobian is not zero or singular. Also it is worth to emphasise again that the coordinate transformations are independent of the action, therefore, the following computations are valid for any symmetric teleparallel theory.

\subsection{Connection set 1}

For the connection set 1 given by \eqref{eq: Gamma set1}, the nontrivial equations \eqref{eq: coincident gauge rewritten} turn out to be 
\begin{subequations}
\begin{align}
    - c \partial_t \xi^\lambda + \partial_t \partial_t \xi^\lambda &=0 \,, \\
- \Gamma^{\phi}{}_{r \phi} \, \partial_t \xi^\lambda + \partial_t \partial_r \xi^\lambda &=0 \,, \\
- c \, \partial_\theta \xi^\lambda + \partial_\theta \partial_t \xi^\lambda &=0 \,, \\
- c \, \partial_\phi \xi^\lambda + \partial_t \partial_\phi \xi^\lambda &=0 \,, \\
- \Gamma^{r}{}_{rr} \, \partial_r \xi^\lambda - \Gamma^{t}{}_{rr} \, \partial_t \xi^\lambda + \partial_r \partial_r \xi^\lambda &=0 \,, \\
- \Gamma^{\phi}{}_{r \phi} \, \partial_\theta \xi^\lambda + \partial_\theta \partial_r \xi^\lambda &=0 \,, \\
- \Gamma^{\phi}{}_{r \phi} \, \partial_\phi \xi^\lambda + \partial_r \partial_\phi \xi^\lambda &=0 \,, \\
\partial_\theta^{2} \xi^\lambda + \frac{1}{c}\partial_t \xi^\lambda &=0 \,, \\
- \cot{\theta} \partial_\phi \xi^\lambda + \partial_\theta \partial_\phi \xi^\lambda &=0 \,, \\
\sin{\theta} \cos{\theta} \, \partial_\theta \xi^\lambda + \partial_\phi \partial_\phi \xi^\lambda + \frac{1}{c}\sin^{2}{\theta} \, \partial_t \xi^\lambda &=0 \,.
\end{align}
\end{subequations}
They are identical for each $\xi^\lambda(t,r,\theta,\phi)$. The general solution of this system of partial differential equations can be written as
\begin{align}
    \xi^\lambda &=  e^{c t + \int \Gamma^{\phi}{}_{r \phi}\, dr} \left[\left(\alpha^\lambda \cos\phi + \beta^\lambda \sin\phi \right) \sin\theta + \gamma^\lambda \cos\theta  \right] + \sigma^\lambda  \int e^{\int \Gamma^{r}{}_{rr}\, dr}\, dr \,, 
    \label{eq: xi} 
\end{align}
where $\alpha^\lambda, \beta^\lambda, \gamma^\lambda, \sigma^\lambda$ are the integration constants. We have kept the indefinite integrals explicitly in the expression \eqref{eq: xi}, so it is straightforward to take the derivatives and check that Eq.\ \eqref{eq: coincident gauge rewritten} is indeed satisfied for the affine connection components \eqref{eq: Gamma set1} and taking into account the relation \eqref{eq: Gamma relations set1}. The solution \eqref{eq: xi} contains only two functions $\Gamma^{\phi}{}_{r \phi}$, $\Gamma^{r}{}_{rr}$, but the third function $\Gamma^{t}{}_{rr}$ would be generated into \eqref{eq: Gamma set1} via the condition \eqref{eq: Gamma relations set1}.

The determinant of the Jacobian matrix is given by
\begin{align}
    \det \left(\frac{\partial \xi^\lambda}{\partial x^\alpha} \right) &= c\,e^{3 c t + \int \Gamma^{\phi}{}_{r \phi}\, dr  + \int \Gamma^{r}{}_{rr}\, dr} \sin \theta \, \det \left( || \alpha^\lambda , \beta^\lambda , \gamma^\lambda , \sigma^\lambda ||
\right)\,,  \label{eq: Jacobian det set1}
\end{align}
where $|| \alpha^\lambda , \beta^\lambda , \gamma^\lambda ,  \sigma^\lambda ||$ stands for a matrix composed of the columns of the integration constants. We see that the coincident gauge coordinates $\xi^\lambda$ are not determined uniquely, but any combination of the integration constants that does not make the Jacobian determinant zero or singular is a possible solution. All these possibilities are of course equivalent to each other in the sense that they are related to each other by some coordinate transformations, as they are all related to the spherical coordinates. Another interesting point is that the coincident coordinates in the static spherical symmetry case depend on the two independent affine connection components allowed by the assumption of zero curvature and torsion. These components remain completely undetermined in STEGR where the connection field equations vanish identically, but could get fixed or partially fixed by the nontrivial connection field equations in extended symmetric teleparallel theories. However the connection contributions appear as exponentiated integrals and are thus guaranteed not to alter the overall sign of the Jacobian determinant and reverse the orientation of the system of coordinates.

A minimal, and reasonable looking choice of integration constants could be $\sigma^0=\alpha^1=\beta^2=\gamma^3=1$ and all the other zero, giving
\begin{subequations}
\label{eq: coinc coordinates set 1}
\begin{align}
\xi^0 &= \xi_w = \int e^{\int \Gamma^{r}{}_{rr}\, dr}\, dr \,, \label{eq: xi_w r set 1} \\
\xi^1 &= \xi_x = e^{c t + \int \Gamma^{\phi}{}_{r \phi}\, dr} \sin{\theta} \cos{\phi} \,, \\
\xi^2 &= \xi_y = e^{c t + \int \Gamma^{\phi}{}_{r \phi}\, dr} \sin{\theta} \sin{\phi} \,, \\
\xi^3 &= \xi_z = e^{c t + \int \Gamma^{\phi}{}_{r \phi}\, dr} \cos{\theta} \,.
\end{align}
\end{subequations}
We may immediately notice that the angular part of this coordinate transformation reminds the relationship between the Cartesian and spherical coordinates, while the time and radial coordinates do not exactly map to their expected Cartesian counterparts. To find that the coincident gauge coordinates resemble the Cartesian system should perhaps not come as a surprise, since in order to make the affine connection to vanish certain ``straightening out'' of the curvilinear nature of the original spherical coordinate system is probably unavoidable. We can invert the expressions \eqref{eq: coinc coordinates set 1} to get
\begin{subequations}
\begin{align}
t &= \frac{1}{2c}\log\Big[R^2 e^{-2 \int \Gamma^\phi{}_{r \phi} dr}\Big] \,, \label{eq: coinc t set 1} \\
r &= r(\xi_w) \,, \\
\theta &= \arccos \left(\frac{\xi_z}{R}\right) \,, \\
\phi &= \arctan \left(\frac{\xi_y}{\xi_x}\right) \,, 
\end{align}
\end{subequations}
where we have denoted
\begin{align}
    R^2 &=\xi_x^2+\xi_y^2+\xi_z^2 \,.
\end{align}
The function $r(\xi_w)$ is the inverse of Eq.\ \eqref{eq: xi_w r set 1} and thus inherits the properties like
\begin{align}
\label{eq: r prime set 1}
    r' &= \frac{dr}{d\xi_w} = e^{\int \Gamma^{r}{}_{rr}\, dr} \,, &
    r'' &= \frac{d^2r}{d\xi_w^2} = - \Gamma^{r}{}_{rr} e^{-2 \int \Gamma^{r}{}_{rr}\, dr} \,,
\end{align}
etc., which must be remembered in the calculations. It is interesting that the time direction $t$ in the spherical coordinates gets mixed into all four coordinates of the coincident gauge, since \eqref{eq: coinc t set 1} depends not only on $\xi_w$ via $\Gamma^\phi{}_{r \phi}(r(\xi_w))$, but also on the ``spatial'' $\xi_x, \xi_y, \xi_z$ via $R$.

To transform the metric and other tensors into the coincident gauge coordinates it is convenient to have the Jacobian matrices written out in these coordinates. Taking the derivatives, keeping in mind the properties \eqref{eq: Gamma relations set1} and \eqref{eq: r prime set 1}, and inverting the matrix, we get
\begin{align}
    \left( \frac{\partial \xi^\lambda}{\partial x^\alpha} \right) &= \begin{pmatrix}0 & \frac{1}{r'} & 0 & 0\\ c\xi_{x} & \xi_{x} \Gamma^{\phi}{}_{r \phi} & \frac{\xi_{x} \xi_{z}}{\sqrt{\xi_{x}^{2} + \xi_{y}^{2}}} & - \xi_{y}\\ c\xi_{y} & \xi_{y} \Gamma^{\phi}{}_{r \phi} & \frac{\xi_{y} \xi_{z}}{\sqrt{\xi_{x}^{2} + \xi_{y}^{2}}} & \xi_{x}\\ c\xi_{z} & \xi_{z} \Gamma^{\phi}{}_{r \phi} & - \sqrt{\xi_{x}^{2} + \xi_{y}^{2}} & 0\end{pmatrix} \,, \\
    \left( \frac{\partial x^\alpha}{\partial \xi^\lambda} \right) &= \begin{pmatrix} \vspace{0.1cm} -\frac{r' \Gamma^{\phi}{}_{r \phi} }{c} & \frac{\xi_{x}}{c R^2} & \frac{\xi_{y}}{c R^2} & \frac{\xi_{z}}{c R^2}\\ r' & 0 & 0 & 0\\0 & \frac{\xi_{x} \xi_{z}}{R^2 \sqrt{\xi_{x}^{2} + \xi_{y}^{2}} } & \frac{\xi_{y} \xi_{z}}{R^2 \sqrt{\xi_{x}^{2} + \xi_{y}^{2}} } & - \frac{\sqrt{\xi_{x}^{2} + \xi_{y}^{2}}}{R^2}\\0 & - \frac{\xi_{y}}{\xi_{x}^{2} + \xi_{y}^{2}} & \frac{\xi_{x}}{\xi_{x}^{2} + \xi_{y}^{2}} & 0\end{pmatrix} \,,
\end{align}
where $\Gamma^{\phi}{}_{r \phi}(r(\xi_w))$, $r=r(\xi_w)$ and $r'=\frac{dr}{d\xi_w}$. With these expressions it is easy to transform \eqref{transf} the metric into the coincident gauge, resulting in
\begin{align}
\label{eq: coinc metric set 1}
    g_{\mu\nu}&=\begin{pmatrix} \vspace{0.1cm} \frac{\left(c^{2} {g_{rr}} - {g_{tt}} \left(\Gamma^{\phi}{}_{r \phi}\right)^{2} \right) r'^{2}}{c^{2}} & \frac{\xi_{x} r' {g_{tt}} \Gamma^{\phi}{}_{r \phi}}{c^{2} R^{2}} & \frac{\xi_{y} r' {g_{tt}} \Gamma^{\phi}{}_{r \phi}}{c^{2} R^{2}} & \frac{\xi_{z} r' {g_{tt}} \Gamma^{\phi}{}_{r \phi}}{c^{2} R^{2}}\\ \vspace{0.1cm} \frac{\xi_{x} r' {g_{tt}} \Gamma^{\phi}{}_{r \phi}}{c^{2} R^{2}} & \frac{ c^{2} \left(\xi_{y}^{2} + \xi_{z}^{2}\right) r^{2} - \xi_{x}^{2} {g_{tt}}}{c^{2} R^{4}} & - \frac{\xi_{x} \xi_{y} \left(c^{2} r^{2} + {g_{tt}}\right)}{c^{2} R^{4}} & - \frac{\xi_{x} \xi_{z} \left(c^{2} r^{2} + {g_{tt}}\right)}{c^{2} R^{4}}\\\vspace{0.1cm} \frac{\xi_{y} r' {g_{tt}} \Gamma^{\phi}{}_{r \phi}}{c^{2} R^{2}} & - \frac{\xi_{x} \xi_{y} \left(c^{2} r^{2} + {g_{tt}}\right)}{c^{2} R^{4}} & \frac{c^{2} \left(\xi_{x}^{2} + \xi_{z}^{2}\right) r^{2}- \xi_{y}^{2} {g_{tt}}}{c^{2} R^{4}} & - \frac{\xi_{y} \xi_{z} \left(c^{2} r^{2} + {g_{tt}}\right)}{c^{2} R^{4}}\\ \vspace{0.1cm} \frac{\xi_{z} r' {g_{tt}} \Gamma^{\phi}{}_{r \phi}}{c^{2} R^{2}} & - \frac{\xi_{x} \xi_{z} \left(c^{2} r^{2} + {g_{tt}}\right)}{c^{2} R^{4}} & - \frac{\xi_{y} \xi_{z} \left(c^{2} r^{2} + {g_{tt}}\right)}{c^{2} R^{4}} & \frac{ c^{2} \left(\xi_{x}^{2} + \xi_{y}^{2}\right) r^{2}- \xi_{z}^{2} {g_{tt}}}{c^{2} R^{4}}\end{pmatrix} \,, 
\end{align}
Similarly, the affine connection components \eqref{eq: Gamma set1} can be transformed \eqref{transf} into the coincident gauge, giving the expected result
\begin{align}
\label{eq: coinc Gamma set 1}
    \Gamma^\rho{}_{\mu \nu}=0 \,.
\end{align}

Note that the transformed metric in the coincident gauge contains now four independent functions, namely $g_{tt}(r(\xi_w))$ and $g_{rr}(r(\xi_w))$ which were in the metric before, as well as $\Gamma^{\phi}{}_{r \phi}(r(\xi_w))$ and $r(\xi_w)$ which previously characterised the connection (the latter via the inverse of \eqref{eq: xi_w r set 1}). Thus although the coordinate transformation into the coincident gauge cleared the affine connection components of all content, the same information got packed into the components of the metric tensor which took a much more complicated form as the result.

Although vanishing connection makes the covariant derivatives to reduce to the partial derivatives and certain simplifications take place, there seems to be no overall economic effect in using the coincident gauge expressions for practical computations. Actually the situation is quite the opposite. Despite zero connection, the nonmetricity tensor is full of nonvanishing components, and thus computing the nonmetricity trances and any further quantities present in the field equations becomes more involved. The reason is not that the all the free functions got packed into the metric, but that the main price of making the connection to vanish was to switch away from the coordinates that were adapted to the symmetry of the configuration, i.e.\ from the spherical coordinates to a sort of deformed Cartesian coordinates. 

The nature of the coincident coordinates \eqref{eq: coinc coordinates set 1} can be illuminated further by transforming the Killing vectors \eqref{eq: Killing spherical set 1} into the coincident gauge, which yields
\begin{subequations}
\label{eq: coinc Killing set 1}
\begin{align}
K^\mu &= \begin{pmatrix} 0 & c \xi_x & c \xi_y & c \xi_z \end{pmatrix} \,, \\
R^\mu &= \begin{pmatrix} 0 & -\xi_y & \xi_x & 0 \end{pmatrix} \,, \\
S^\mu &= \begin{pmatrix} 0 & \xi_z & 0 & -\xi_x \end{pmatrix} \,, \\
T^\mu &= \begin{pmatrix} 0 & 0 & -\xi_z & \xi_y \end{pmatrix} \,.
\end{align}
\end{subequations}
In the form of the vectors $R^\mu$, $S^\mu$, $T^\mu$ we can exactly recognize the Killing vectors of spherical symmetry as expressed in the Cartesian coordinates, but the vector $K^\mu$ representing static time translation symmetry is quite unexpectedly orthogonal to the $\xi_w$ direction. Therefore we can not entertain $\xi_w$ as a coincident gauge analogue of time, rather the time is ``smeared'' among all the coordinates. Hence one should be alert that in general the physical interpretation of solutions in the coincident gauge can be tricky and problematic.
However, one can still check explicitly that the the Lie derivatives \eqref{LieD_mag} of the metric \eqref{eq: coinc metric set 1} in the directions specified by the Killing vectors \eqref{eq: coinc Killing set 1} vanish, hence the transformed metric \eqref{eq: coinc metric set 1} possesses the same symmetry as the original metric \eqref{eq: spherical metric} and connection \eqref{eq: Gamma set1}. In a similar vein, the Lie derivatives of the transformed connection components \eqref{eq: coinc Gamma set 1} in the directions specified by the Killing vectors \eqref{eq: coinc Killing set 1} vanish as well. 

Taking the Minkowski limit of the original metric by letting $g_{tt}\to 1$ and $g_{rr}\to 1$ simplifies the form of of the coincident gauge metric \eqref{eq: coinc metric set 1}. However, by closer inspection we may notice that no possible choice of the remaining functions $\Gamma^{\phi}{}_{r \phi}$ and $\Gamma^{r}{}_{rr}$ (the latter determines $r$) can take the Minkowski limit of \eqref{eq: coinc metric set 1} into the familiar form in the Cartesian coordinates $\eta_{\mu\nu}=\textrm{diag}(-1,1,1,1)$. At this point it is not obvious that the  Minkowski limit of the coincident gauge metric should reduce to the Cartesian coordinates, but on the other hand the Minkowski spacetime in the Cartesian coordinates is already trivially in the coincident gauge, and it feels unsettling that the set 1 connection does not possess a direct link to that.

\subsection{Connection set 2 general case (branch 1)}
For the connection set 2 expressed by \eqref{eq: Gamma set2}, the nontrivial equations \eqref{eq: coincident gauge rewritten} turn out to be 
\begin{subequations}
\label{eq: set 2 branch 1 system}
\begin{align}
c \left(2 c - k\right) \Gamma^{r}{}_{\theta \theta} \,  \partial_r \xi^\lambda - \left(- c \left(2 c - k\right) \Gamma^{t}{}_{\theta \theta} \,  - c + k\right) \partial_t \xi^\lambda + \partial_t \partial_t \xi^\lambda &=0 \,, \\
- \frac{\left(2 c - k\right) \left(c \Gamma^{t}{}_{\theta \theta} \,  + 1\right) \Gamma^{t}{}_{\theta \theta} \,  \partial_t \xi^\lambda}{\Gamma^{r}{}_{\theta \theta} \, } - \left(c \left(2 c - k\right) \Gamma^{t}{}_{\theta \theta} \,  + c\right) \partial_r \xi^\lambda + \partial_t\partial_r \xi^\lambda &=0 \,, \\
- c \, \partial_\theta \xi^\lambda + \partial_\theta\partial_t \xi^\lambda &=0 \,, \\
- c \, \partial_\phi \xi^\lambda + \partial_t\partial_\phi \xi^\lambda &=0 \,, \\
- \Gamma^{r}{}_{rr} \,  \partial_r \xi^\lambda - \Gamma^{t}{}_{rr} \,  \partial_t \xi^\lambda + \partial_r \partial_r \xi^\lambda &=0 \,, \\
- \frac{\left(- c \Gamma^{t}{}_{\theta \theta} \,  - 1\right) \partial_\theta \xi^\lambda}{\Gamma^{r}{}_{\theta \theta} \, } + \partial_\theta \partial_r \xi^\lambda &=0 \,, \\
- \frac{\left(- c \Gamma^{t}{}_{\theta \theta} \,  - 1\right) \partial_\phi \xi^\lambda}{\Gamma^{r}{}_{\theta \theta} \, } + \partial_r \partial_\phi \xi^\lambda &=0 \,, \\
- \Gamma^{r}{}_{\theta \theta} \,  \partial_r \xi^\lambda - \Gamma^{t}{}_{\theta \theta} \,  \partial_t \xi^\lambda + \partial_\theta \partial_\theta \xi^\lambda &=0 \,, \\
- \cot{\theta} \, \partial_\phi \xi^\lambda + \partial_\theta \partial_\phi \xi^\lambda &=0 \,, \\
- \Gamma^{r}{}_{\theta \theta} \,  \sin^{2}{\theta} \, \partial_r \xi^\lambda - \Gamma^{t}{}_{\theta \theta} \,  \sin^{2}{\theta} \, \partial_t \xi^\lambda + \sin{\theta} \cos{\theta} \, \partial_\theta \xi^\lambda + \partial_\phi \partial_\phi \xi^\lambda &=0 \,,
\end{align}
\end{subequations}
where again, the remaining equations are identical for each $\xi^\lambda(t,r,\theta,\phi)$. Assuming $c \neq k \neq 0$, the general solution of the system is 
\begin{align}
    \xi^\lambda &=
     e^{c t - \int \frac{c \Gamma^{t}{}_{\theta \theta} + 1}{\Gamma^{r}{}_{\theta \theta}}\, dr}\left(\left(\alpha^\lambda \cos\phi + \beta^\lambda \sin\phi \right) \sin\theta + \gamma^\lambda \cos\theta  \right)
     + \sigma^\lambda e^{(k-c) (t - \int \frac{\Gamma^{t}{}_{\theta \theta}}{\Gamma^{r}{}_{\theta \theta}}\, dr )} \,.
     \label{eq: xi set2B}
\end{align}
Thus the coordinate system of the coincident gauge can take various forms depending on the choice of the integration constants $\alpha^\lambda, \beta^\lambda, \gamma^\lambda, \sigma^\lambda$. All these possibilities are permissible, provided that the Jacobian determinant
\begin{align}
    \det \left(\frac{\partial \xi^\lambda}{\partial x^\alpha} \right) &= \frac{(c-k) e^{3\left(ct - \int \frac{c\Gamma^{t}{}_{\theta \theta}+1}{\Gamma^{r}{}_{\theta \theta}}\, dr \right)-(k-c)\left(t - b\int \frac{\Gamma^{t}{}_{\theta \theta}}{\Gamma^{r}{}_{\theta \theta}}\, dr \right) } \sin \theta}{\Gamma^{r}{}_{\theta \theta}}  \det \left( || \alpha^\lambda , \beta^\lambda , \gamma^\lambda , \sigma^\lambda ||
\right)\,,  \label{eq: Jacobian det set2}
\end{align}
is regular. The expression \eqref{eq: xi set2B} contains only the functions $\Gamma^{t}{}_{\theta \theta}$ and $\Gamma^{r}{}_{\theta \theta}$ while the other two functions $\Gamma^{t}{}_{rr}$ and $\Gamma^{r}{}_{rr}$ in the connection
\eqref{eq: Gamma set2} are generated by the conditions \eqref{eq: Gamma relations set2}.

A minimalistic choice would be to take $\sigma^0=\alpha^1=\beta^2=\gamma^3=1$ and let the remaining constants to vanish. This gives
\begin{subequations}
\label{eq: coinc coordinates set 2}
\begin{align}
\xi^0 &= \xi_w = e^{(k-c) \left(t - \int \frac{\Gamma^{t}{}_{rr}}{\Gamma^{r}{}_{rr}}\, dr \right) } \,, \label{eq: xi_w r set 22B} \\
\xi^1 &= \xi_x = e^{c t - \int \frac{c \Gamma^{t}{}_{\theta \theta} + 1}{\Gamma^{r}{}_{\theta \theta}}\, dr} \sin{\theta} \cos{\phi} \,, \\
\xi^2 &= \xi_y = e^{c t - \int \frac{c \Gamma^{t}{}_{\theta \theta} + 1}{\Gamma^{r}{}_{\theta \theta}}\, dr} \sin{\theta} \sin{\phi} \,, \\
\xi^3 &= \xi_z = e^{c t - \int \frac{c \Gamma^{t}{}_{\theta \theta} + 1}{\Gamma^{r}{}_{\theta \theta}}\, dr} \cos{\theta} \,.
\end{align}
\end{subequations}
We can invert the expressions \eqref{eq: coinc coordinates set 2} to get
\begin{subequations}
\begin{align}
t &= \frac{1}{c}\log\Big[R e^{\int \frac{c \Gamma^t{}_{\theta \theta} +1}{\Gamma^r{}_{\theta\theta}} dr}\Big]  = \frac{1}{c-k} \log \Big[\xi_w \, e^{(c-k)\int \frac{\Gamma^t{}_{rr}}{\Gamma^r{}_{rr}} dr} \Big] \,, \label{eq: coinc t set 2}\\
r &= r(\xi_w,\xi_x,\xi_y,\xi_z) \,, \\
\theta &= \arccos \left(\frac{\xi_z}{R}\right) \,, \\
\phi &= \arctan \left(\frac{\xi_y}{\xi_x}\right) \,, 
\end{align}
\end{subequations}
where we have denoted
\begin{align}
    R^2 &=\xi_x^2+\xi_y^2+\xi_z^2 \,.
\end{align}
The function $r(\xi_w,\xi_x,\xi_y,\xi_z)$ is not arbitrary, but is obtained by inverting Eqs.\ \eqref{eq: coinc coordinates set 2} and thus inherits several relations for its derivatives.

To proceed with transformations into the coincident gauge coordinates it is again convenient to have the Jacobian matrices written out in these coordinates. Taking the derivatives, we get
\begin{align}
    \left( \frac{\partial \xi^\lambda}{\partial x^\alpha} \right) &= \begin{pmatrix}\xi_{w} \left(k-c\right) & \frac{\xi_{w} \left(c - k\right) \Gamma^{t}{}_{\theta \theta}}{\Gamma^{r}{}_{\theta \theta}} & 0 & 0\\\xi_{x} c & - \frac{\xi_{x} \left(c \Gamma^{t}{}_{\theta \theta} + 1\right)}{\Gamma^{r}{}_{\theta \theta}} & \frac{\xi_{x} \xi_{z}}{\sqrt{\xi_{x}^{2} + \xi_{y}^{2}}} & - \xi_{y}\\\xi_{y} c & - \frac{\xi_{y} \left(c \Gamma^{t}{}_{\theta \theta} + 1\right)}{\Gamma^{r}{}_{\theta \theta}} & \frac{\xi_{y} \xi_{z}}{\sqrt{\xi_{x}^{2} + \xi_{y}^{2}}} & \xi_{x}\\\xi_{z} c & - \frac{\xi_{z} \left(c \Gamma^{t}{}_{\theta \theta} + 1\right)}{\Gamma^{r}{}_{\theta \theta}} & - \sqrt{\xi_{x}^{2} + \xi_{y}^{2}} & 0\end{pmatrix} 
    \,, \\
    \left( \frac{\partial x^\alpha}{\partial \xi^\lambda} \right) &= \begin{pmatrix} \vspace{0.1cm} \frac{c \Gamma^{t}{}_{\theta \theta} + 1}{\xi_{w} \left(k-c\right)} & -\frac{\xi_{x} \Gamma^{t}{}_{\theta \theta}}{R^{2}} & - \frac{\xi_{y} \Gamma^{t}{}_{\theta \theta}}{R^{2}} & - \frac{\xi_{z} \Gamma^{t}{}_{\theta \theta}}{R^{2}}\\ \vspace{0.1cm} \frac{c \Gamma^{r}{}_{\theta \theta}}{\xi_{w} \left(k-c\right)} & - \frac{\xi_{x} \Gamma^{r}{}_{\theta \theta}}{R^{2}} & - \frac{\xi_{y} \Gamma^{r}{}_{\theta \theta}}{R^{2}} & - \frac{\xi_{z} \Gamma^{r}{}_{\theta \theta}}{R^{2}}\\ \vspace{0.1cm} 0 & \frac{\xi_{x} \xi_{z}}{R^{2} \sqrt{\xi_{x}^{2} + \xi_{y}^{2}}} & \frac{\xi_{y} \xi_{z}}{R^{2} \sqrt{\xi_{x}^{2} + \xi_{y}^{2}}} & - \frac{\sqrt{\xi_{x}^{2} + \xi_{y}^{2}}}{R^{2}}\\0 & - \frac{\xi_{y}}{\xi_{x}^{2} + \xi_{y}^{2}} & \frac{\xi_{x}}{\xi_{x}^{2} + \xi_{y}^{2}} & 0\end{pmatrix} \,,
\end{align}
where $\Gamma^{t}{}_{\theta\theta}$, and $\Gamma^{r}{}_{\theta\theta}$ are functions of  $r(\xi_w,\xi_x,\xi_y,\xi_z)$. With these expressions it is straightforward to transform \eqref{transf} the metric \eqref{eq: spherical metric}
into the coincident gauge as
\begin{align}
    g_{\mu\nu}&=\left( \begin{matrix} \vspace{0.1cm} \frac{c^{2} \left(\Gamma^{r}{}_{\theta \theta}\right)^{2} g_{rr} - c^{2} \left(\Gamma^{t}{}_{\theta \theta}\right)^{2} g_{tt} - 2 c \Gamma^{t}{}_{\theta \theta} g_{tt} - g_{tt}}{\xi_{w}^{2} \left(c - k\right)^{2}} & \frac{\xi_{x} \left(c \left(\Gamma^{r}{}_{\theta \theta}\right)^{2} g_{rr} - c \left(\Gamma^{t}{}_{\theta \theta}\right)^{2} g_{tt} - \Gamma^{t}{}_{\theta \theta} g_{tt}\right)}{R^{2} \xi_{w} \left(c - k\right)} 
    \\ \vspace{0.1cm} \frac{\xi_{x} \left(c \left(\Gamma^{r}{}_{\theta \theta}\right)^{2} g_{rr} - c \left(\Gamma^{t}{}_{\theta \theta}\right)^{2} g_{tt} - \Gamma^{t}{}_{\theta \theta} g_{tt}\right)}{R^{2} \xi_{w} \left(c - k\right)} & \frac{\xi_{x}^{2} \left(\Gamma^{r}{}_{\theta \theta}\right)^{2} g_{rr} - \xi_{x}^{2} \left(\Gamma^{t}{}_{\theta \theta}\right)^{2} g_{tt} + \xi_{y}^{2} r^{2} + \xi_{z}^{2} r^{2}}{R^{4}} 
    \\ \vspace{0.1cm} \frac{\xi_{y} \left(c \left(\Gamma^{r}{}_{\theta \theta}\right)^{2} g_{rr} - c \left(\Gamma^{t}{}_{\theta \theta}\right)^{2} g_{tt} - \Gamma^{t}{}_{\theta \theta} g_{tt}\right)}{R^{2} \xi_{w} \left(c - k\right)} & - \frac{\xi_{x} \xi_{y} \left(r^{2} - \left(\Gamma^{r}{}_{\theta \theta}\right)^{2} g_{rr} + \left(\Gamma^{t}{}_{\theta \theta}\right)^{2} g_{tt}\right)}{R^{4}} 
    \\ \vspace{0.1cm} \frac{\xi_{z} \left(c \left(\Gamma^{r}{}_{\theta \theta}\right)^{2} g_{rr} - c \left(\Gamma^{t}{}_{\theta \theta}\right)^{2} g_{tt} - \Gamma^{t}{}_{\theta \theta} g_{tt}\right)}{R^{2} \xi_{w} \left(c - k\right)} & - \frac{\xi_{x} \xi_{z} \left(r^{2} - \left(\Gamma^{r}{}_{\theta \theta}\right)^{2} g_{rr} + \left(\Gamma^{t}{}_{\theta \theta}\right)^{2} g_{tt}\right)}{R^{4}}  \end{matrix} \right.
    \nonumber \\
    & \qquad \qquad \left. \begin{matrix} \vspace{0.1cm}  & \frac{\xi_{y} \left(c \left(\Gamma^{r}{}_{\theta \theta}\right)^{2} g_{rr} - c \left(\Gamma^{t}{}_{\theta \theta}\right)^{2} g_{tt} - \Gamma^{t}{}_{\theta \theta} g_{tt}\right)}{R^{2} \xi_{w} \left(c - k\right)} & \frac{\xi_{z} \left(c \left(\Gamma^{r}{}_{\theta \theta}\right)^{2} g_{rr} - c \left(\Gamma^{t}{}_{\theta \theta}\right)^{2} g_{tt} - \Gamma^{t}{}_{\theta \theta} g_{tt}\right)}{R^{2} \xi_{w} \left(c - k\right)} \\
    \vspace{0.1cm} & - \frac{\xi_{x} \xi_{y} \left(r^{2} - \left(\Gamma^{r}{}_{\theta \theta}\right)^{2} g_{rr} + \left(\Gamma^{t}{}_{\theta \theta}\right)^{2} g_{tt}\right)}{R^{4}} & - \frac{\xi_{x} \xi_{z} \left(r^{2} - \left(\Gamma^{r}{}_{\theta \theta}\right)^{2} g_{rr} + \left(\Gamma^{t}{}_{\theta \theta}\right)^{2} g_{tt}\right)}{R^{4}} \\
    \vspace{0.1cm} & \frac{\xi_{x}^{2} r^{2} + \xi_{y}^{2} \left(\Gamma^{r}{}_{\theta \theta}\right)^{2} g_{rr} - \xi_{y}^{2} \left(\Gamma^{t}{}_{\theta \theta}\right)^{2} g_{tt} + \xi_{z}^{2} r^{2}}{R^{4}} & - \frac{\xi_{y} \xi_{z} \left(r^{2} - \left(\Gamma^{r}{}_{\theta \theta}\right)^{2} g_{rr} + \left(\Gamma^{t}{}_{\theta \theta}\right)^{2} g_{tt}\right)}{R^{4}} \\
    & - \frac{\xi_{y} \xi_{z} \left(r^{2} - \left(\Gamma^{r}{}_{\theta \theta}\right)^{2} g_{rr} + \left(\Gamma^{t}{}_{\theta \theta}\right)^{2} g_{tt}\right)}{R^{4}} & \frac{\xi_{x}^{2} r^{2} + \xi_{y}^{2} r^{2} + \xi_{z}^{2} \left(\Gamma^{r}{}_{\theta \theta}\right)^{2} g_{rr} - \xi_{z}^{2} \left(\Gamma^{t}{}_{\theta \theta}\right)^{2} g_{tt}}{R^{4}} 
    \end{matrix} \right)
\label{eq: coinc metric set 2}
\end{align}
Similarly, the affine connection components \eqref{eq: Gamma set2} can be transformed into the coincident gauge, resulting in the desired expression
\begin{align}
\label{eq: coinc Gamma set 2}
    \Gamma^\lambda{}_{\mu \nu}=0 \,.
\end{align} 

The nature of the coincident coordinates is quite similar to the case of connection set 1, namely the angular variables map into the typical Cartesian equivalents, but the time and radial coordinate get mixed among all the directions. This property is illustrated by the Killing vectors \eqref{eq: Killing spherical set 1} transformed into the coincident gauge as
\begin{subequations}
\label{eq: coinc Killing set 2}
\begin{align}
K^\mu &= \begin{pmatrix} (k-c) \xi_w & c \xi_x & c \xi_y & c \xi_z \end{pmatrix} \,, \\
R^\mu &= \begin{pmatrix} 0 & -\xi_y & \xi_x & 0 \end{pmatrix} \,, \\
S^\mu &= \begin{pmatrix} 0 & \xi_z & 0 & -\xi_x \end{pmatrix} \,, \\
T^\mu &= \begin{pmatrix} 0 & 0 & -\xi_z & \xi_y \end{pmatrix} \,.
\end{align}
\end{subequations}
The form of the vectors $R^\mu$, $S^\mu$, $T^\mu$ corresponds exactly to the Killing vectors of spherical symmetry as expressed in the Cartesian coordinates, but the vector $K^\mu$ representing static time translation symmetry has strangely nonzero compinents in all directions $\xi^\lambda$. Similarly to the set 1 case, in the Minkowski limit the metric \eqref{eq: coinc metric set 2} can not be reduced to the Cartesian form.

\subsection{Connection set 2 special case (branch 2)}

When the the parameters $c$ and $k$ are zero, the qualitative features of the system \eqref{eq: set 2 branch 1 system} change and the solution \eqref{eq: xi set2B} does not give a correct coordinate transformation since the Jacobian determinant \eqref{eq: Jacobian det set2} vanishes. However, the case $c=k=0$ is important as it leads to nontrivial balck hole solutions in $f(Q)$ \cite{DAmbrosio:2021zpm} and scalar-tensor \cite{Bahamonde:2022esv} symmetric teleparallel gravity. Therefore we need to tackle the equations \eqref{eq: coincident gauge rewritten} for the connection set 2 (see Eq.~\eqref{eq: Gamma set2}) with $c=k=0$ separately. The system of nontrivial equations turns out to be 
\begin{subequations}
\begin{align}
\partial_t \partial_t \xi^\lambda &=0 \,, \\
\partial_t\partial_r \xi^\lambda &=0 \,, \\
\partial_\theta \partial_t \xi^\lambda &=0 \,, \\
\partial_t \partial_\phi \xi^\lambda &=0 \,, \\
- \Gamma^{r}{}_{rr} \,  \partial_r \xi^\lambda - \Gamma^{t}{}_{rr} \,  \partial_t \xi^\lambda + \partial_r \partial_r \xi^\lambda &=0 \,, \\
\partial_\theta \partial_r \xi^\lambda + \frac{\partial_\theta \xi^\lambda}{\Gamma^{r}{}_{\theta \theta}} &=0 \,, \\
\partial_r\partial_\phi \xi^\lambda + \frac{\partial_\phi \xi^\lambda}{\Gamma^{r}{}_{\theta \theta}} &=0 \,, \\
- \Gamma^{r}{}_{\theta \theta} \, \partial_r \xi^\lambda - \Gamma^{t}{}_{\theta \theta} \,  \partial_t \xi^\lambda + \partial_\theta \partial_\theta \xi^\lambda &=0 \,, \\
- \cot{\theta} \, \partial_\phi \xi^\lambda + \partial_\theta \partial_\phi \xi^\lambda &=0 \,, \\
- \Gamma^{r}{}_{\theta \theta} \, \sin^{2}{\theta} \, \partial_r \xi^\lambda - \Gamma^{t}{}_{\theta \theta} \, \sin^{2}{\theta} \, \partial_t \xi^\lambda + \sin{\theta} \cos{\theta} \, \partial_\theta \xi^\lambda + \partial_\phi \partial_\phi \xi^\lambda &=0 \,. 
\end{align}
\end{subequations}
The general solution is given by
\begin{align}
    \xi^\lambda &=  e^{-\int \frac{1}{\Gamma^{r}{}_{\theta\theta}} \, dr} \left[\left(\alpha^\lambda \cos(\phi) + \beta^\lambda \sin(\phi) \right) \sin(\theta ) + \gamma^\lambda \cos(\theta)  \right] + \sigma^\lambda \left( t - \int \frac{\Gamma^{t}{}_{\theta\theta}}{\Gamma^{r}{}_{\theta\theta}}\, dr \right)  \,.
    \label{eq: xi set 22} 
\end{align}
Note that there are some qualitative differences with the previous case expression \eqref{eq: xi set2B}, and the latter does not reduce to the solution \eqref{eq: xi set 22} above by setting $c=k=0$ in \eqref{eq: xi set2B}. Nevertheless the overall structure of the present solutions is similar with four integration constants and integrals of the free functions. By taking the derivatives one can check explicitly that the connection \eqref{eq: Gamma set2} and \eqref{eq: xi set 22} satisfy \eqref{eq: coincident gauge rewritten}. The two other functions $\Gamma^{t}{}_{rr}$ and $\Gamma^{r}{}_{rr}$ are taken care of by the relations \eqref{eq: Gamma relations set2}
\begin{subequations}
\label{eq: Gamma relations set22}
\begin{align}
    \frac{d}{d r} \Gamma^{t}{}_{\theta \theta}  &= - \frac{\Gamma^{t}{}_{\theta \theta} }{\Gamma^{r}{}_{\theta \theta} } - \Gamma^{r}{}_{\theta \theta}  \Gamma^{t}{}_{rr} \,, \\
    \frac{d}{d r} \Gamma^{r}{}_{\theta \theta}  &= - \Gamma^{r}{}_{\theta \theta}  \Gamma^{r}{}_{rr}  - 1 \,.
\end{align}
\end{subequations}

Again by picking different nonzero values of the integration constants the coincident gauge coordinate system will take different forms. All these are valid solutions as long as the Jacobian determinant 
\begin{align}
    \det \left(\frac{\partial \xi^\lambda}{\partial x^\alpha} \right) &= \frac{e^{ - 3 \int \frac{1}{\Gamma^{r}{}_{\theta \theta} }\, dr} \sin \theta }{\Gamma^{r}{}_{\theta \theta}}  \det \left( || \alpha^\lambda , \beta^\lambda , \gamma^\lambda , \sigma^\lambda ||
\right)\,,  \label{eq: Jacobian det set2B}
\end{align}
is regular and non-zero.
As we did in the previous sections, a simple choice would be to take $\sigma^0=\alpha^1=\beta^2=\gamma^3=1$ and let the remaining constants to vanish. This gives
\begin{subequations}
\label{eq: coinc coordinates set 22}
\begin{align}
\xi^0 &= \xi_w = t - \int \frac{\Gamma^{t}{}_{\theta\theta}}{\Gamma^{r}{}_{\theta\theta}}\, dr \,, \label{eq: xi_w r set 22} \\
\xi^1 &= \xi_x = e^{ - \int \frac{1}{\Gamma^{r}{}_{\theta \theta}}\, dr} \sin{\theta} \cos{\phi} \,, \\
\xi^2 &= \xi_y = e^{- \int \frac{1}{\Gamma^{r}{}_{\theta \theta}}\, dr} \sin{\theta} \sin{\phi} \,, \\
\xi^3 &= \xi_z = e^{- \int \frac{1}{\Gamma^{r}{}_{\theta \theta}}\, dr} \cos{\theta} \,.
\end{align}
\end{subequations}
 We can invert these expressions \eqref{eq: coinc coordinates set 22} to get
\begin{subequations}
\begin{align}
t &= \xi_w + \int \frac{\Gamma^{t}{}_{\theta\theta}}{\Gamma^{r}{}_{\theta\theta}}\, dr  \,, \label{eq: coinc t set 22} \\
r &= r(R) \,, \\
\theta &= \arccos \left(\frac{\xi_z}{R}\right) \,, \\
\phi &= \arctan \left(\frac{\xi_y}{\xi_x}\right) \,, 
\end{align}
\end{subequations}
where we have denoted
\begin{align}
    R^2 &=\xi_x^2+\xi_y^2+\xi_z^2 \,.
\end{align}
The function $r(R(\xi_x, \xi_y, \xi_z))$ is the inverse of the last three equations of \eqref{eq: coinc coordinates set 22} and thus inherits the properties like
\begin{align}
\label{eq: r prime set 22}
    \frac{dr}{dR} &= -\frac{\Gamma^{r}{}_{\theta\theta}}{R} \,, \qquad 
\frac{dR}{d\xi_i} = \frac{\xi_i}{R} \,,
\end{align}
and chain rule must be taken into account in the calculations.

The Jacobian matrix and its inverse in the coincident coordinates \eqref{eq: coinc coordinates set 22} are
\begin{align}
    \left( \frac{\partial \xi^\lambda}{\partial x^\alpha} \right) &= \begin{pmatrix}\vspace{0.1cm}1 & - \frac{\Gamma^{t}{}_{\theta \theta}}{\Gamma^{r}{}_{\theta \theta}} & 0 & 0\\\vspace{0.1cm}0 & - \frac{\xi_{x}}{\Gamma^{r}{}_{\theta \theta}} & \frac{\xi_{x} \xi_{z}}{\sqrt{\xi_{x}^{2} + \xi_{y}^{2}}} & - \xi_{y}\\\vspace{0.1cm}0 & - \frac{\xi_{y}}{\Gamma^{r}{}_{\theta \theta}} & \frac{\xi_{y} \xi_{z}}{\sqrt{\xi_{x}^{2} + \xi_{y}^{2}}} & \xi_{x}\\0 & - \frac{\xi_{z}}{\Gamma^{r}{}_{\theta \theta}} & - \sqrt{\xi_{x}^{2} + \xi_{y}^{2}} & 0\end{pmatrix} \,, 
    \\
    \left( \frac{\partial x^\lambda}{\partial \xi^\alpha} \right) &= \begin{pmatrix}\vspace{0.1cm}1 & - \frac{\xi_{x} \Gamma^{t}{}_{\theta \theta}}{R^{2}} & - \frac{\xi_{y} \Gamma^{t}{}_{\theta \theta}}{R^{2}} & - \frac{\xi_{z} \Gamma^{t}{}_{\theta \theta}}{R^{2}}\\\vspace{0.1cm}0 & - \frac{\xi_{x} \Gamma^{r}{}_{\theta \theta}}{R^{2}} & - \frac{\xi_{y} \Gamma^{r}{}_{\theta \theta}}{R^{2}} & - \frac{\xi_{z} \Gamma^{r}{}_{\theta \theta}}{R^{2}}\\\vspace{0.1cm}0 & \frac{\xi_{x} \xi_{z}}{R^{2} \sqrt{\xi_{x}^{2} + \xi_{y}^{2}}} & \frac{\xi_{y} \xi_{z}}{R^{2} \sqrt{\xi_{x}^{2} + \xi_{y}^{2}}} & - \frac{\sqrt{\xi_{x}^{2} + \xi_{y}^{2}}}{R^{2}}\\0 & - \frac{\xi_{y}}{\xi_{x}^{2} + \xi_{y}^{2}} & \frac{\xi_{x}}{\xi_{x}^{2} + \xi_{y}^{2}} & 0\end{pmatrix} \,.
\end{align}
This helps us to transform \eqref{transf} the metric \eqref{eq: spherical metric} into the coincident gauge as
\begin{align}
    g_{\mu\nu}&= \left( \begin{matrix} - g_{tt} & \frac{\xi_{x} \Gamma^{t}{}_{\theta \theta} g_{tt}}{R^{2}} & \frac{\xi_{y} \Gamma^{t}{}_{\theta \theta} g_{tt}}{R^{2}} & 
    \\
    \frac{\xi_{x} \Gamma^{t}{}_{\theta \theta} g_{tt}}{R^{2}} & \frac{\xi_{x}^{2} \left(\Gamma^{r}{}_{\theta \theta}\right)^{2} g_{rr} - \xi_{x}^{2} \left(\Gamma^{t}{}_{\theta \theta}\right)^{2} g_{tt} + \xi_{y}^{2} r^{2} + \xi_{z}^{2} r^{2}}{R^{4}} & - \frac{\xi_{x} \xi_{y} \left(r^{2} - \left(\Gamma^{r}{}_{\theta \theta}\right)^{2} g_{rr} + \left(\Gamma^{t}{}_{\theta \theta}\right)^{2} g_{tt}\right)}{R^{4}} & \hspace{1cm}
    \\
    \frac{\xi_{y} \Gamma^{t}{}_{\theta \theta} g_{tt}}{R^{2}} & - \frac{\xi_{x} \xi_{y} \left(r^{2} - \left(\Gamma^{r}{}_{\theta \theta}\right)^{2} g_{rr} + \left(\Gamma^{t}{}_{\theta \theta}\right)^{2} g_{tt}\right)}{R^{4}} & \frac{\xi_{x}^{2} r^{2} + \xi_{y}^{2} \left(\Gamma^{r}{}_{\theta \theta}\right)^{2} g_{rr} - \xi_{y}^{2} \left(\Gamma^{t}{}_{\theta \theta}\right)^{2} g_{tt} + \xi_{z}^{2} r^{2}}{R^{4}} &
    \\
    \frac{\xi_{z} \Gamma^{t}{}_{\theta \theta} g_{tt}}{R^{2}} & - \frac{\xi_{x} \xi_{z} \left(r^{2} - \left(\Gamma^{r}{}_{\theta \theta}\right)^{2} g_{rr} + \left(\Gamma^{t}{}_{\theta \theta}\right)^{2} g_{tt}\right)}{R^{4}} & - \frac{\xi_{y} \xi_{z} \left(r^{2} - \left(\Gamma^{r}{}_{\theta \theta}\right)^{2} g_{rr} + \left(\Gamma^{t}{}_{\theta \theta}\right)^{2} g_{tt}\right)}{R^{4}} & 
 \end{matrix} 
 \right. \nonumber \\
    & \hspace{8cm} \left. 
    \begin{matrix} 
    \frac{\xi_{z} \Gamma^{t}{}_{\theta \theta} g_{tt}}{R^{2}} \\ - \frac{\xi_{x} \xi_{z} \left(r^{2} - \left(\Gamma^{r}{}_{\theta \theta}\right)^{2} g_{rr} + \left(\Gamma^{t}{}_{\theta \theta}\right)^{2} g_{tt}\right)}{R^{4}} \\  - \frac{\xi_{y} \xi_{z} \left(r^{2} - \left(\Gamma^{r}{}_{\theta \theta}\right)^{2} g_{rr} + \left(\Gamma^{t}{}_{\theta \theta}\right)^{2} g_{tt}\right)}{R^{4}} \\ \frac{\xi_{x}^{2} r^{2} + \xi_{y}^{2} r^{2} + \xi_{z}^{2} \left(\Gamma^{r}{}_{\theta \theta}\right)^{2} g_{rr} - \xi_{z}^{2} \left(\Gamma^{t}{}_{\theta \theta}\right)^{2} g_{tt}}{R^{4}} \end{matrix} \right) \,,
\label{eq: coinc metric set 22}
\end{align}
while the transformed \eqref{transf} connection components \eqref{eq: Gamma set2} with $c=k=0$ are of course
\begin{align}
    \Gamma^\rho{}_{\mu\nu} &=0 \,.
\end{align}

The nature of the coincident gauge coordinates is clarified by the Killing vectors \eqref{eq: Killing spherical set 1} which get transformed into
\begin{subequations}
\label{eq: coinc Killing set 22}
\begin{align}
K^\mu &= \begin{pmatrix} 1 & 0 & 0 & 0 \end{pmatrix} \,, \\
R^\mu &= \begin{pmatrix} 0 & -\xi_y & \xi_x & 0 \end{pmatrix} \,, \\
S^\mu &= \begin{pmatrix} 0 & \xi_z & 0 & -\xi_x \end{pmatrix} \,, \\
T^\mu &= \begin{pmatrix} 0 & 0 & -\xi_z & \xi_y \end{pmatrix} \,.
\end{align}
\end{subequations}
Here we see that not only do the Killing vectors of the spherical symmetry exactly take the form as in the Cartesian coordinates, but also the time translation vector $K^\mu$ responsible for staticity appears in the form as in the Cartesian coordinates. This lands us on a more familiar ground when the interpretation of some field configuration in the coincident gauge is needed. Furthermore, in the Minkowski limit $g_{tt} \to 1$, $g_{rr} \to 1$ this metric reduces to the familiar $\eta_{\mu\nu}$ of Cartesian coordinates if $\Gamma^{t}{}_{\theta \theta} \to 0$, $\Gamma^{r}{}_{\theta \theta} \to \pm r$. 

\section{Minkowski limit}\label{sec:Mink}

Noticing the issues arising with the Minkowski limit in the coincident gauge motivates us to check how do the connections \eqref{eq: Gamma set1} and \eqref{eq: Gamma set2} behave in the Minkowski limit 
\begin{align}
\label{eq: Minkowski metric}
    g_{tt} & = 1 \,, \qquad g_{rr} = 1 \,,
\end{align}
already in the spherical coordinates. In this limit all effects of gravity and inertia should vanish. One would expect that as this happens the nonmetricity tensor and nonmetricity scalar should also vanish identically, as there should be no residual nontrivial geometry left. As reasonable as this assumption sounds, it might not immediately meet an unanimous approval. The reason is that as matter is introduced into the theory by a minimal or metric coupling principle \cite{BeltranJimenez:2020sih}, it follows the Levi-Civita geodesics in symmetric teleparallel gravity. In other words, the particles only get to feel the properties of the metric, and as long as the metric is Minkowski, nontrivial nonmetricity in the background is impossible to detect directly. We will elaborate on this point some more in the discussion Sec.~\ref{sec:conclusions}, and proceed here by calculating the nonmetricity tensor and scalar in the different sets.

\subsection{Connection set 1}

In spherical coordinates, taking the metric \eqref{eq: spherical metric} to be Minkowski \eqref{eq: Minkowski metric}, and the connection to be given by Eq.\ \eqref{eq: Gamma set1}, the nonmetricity tensor \eqref{NonMetricityTensor} is given by
\small{\begin{align}
    Q_{\rho\mu\nu} &=\left[\begin{matrix}\left[\begin{matrix}2 c & \Gamma^{\phi}{}_{r \phi} & 0 & 0\\\Gamma^{\phi}{}_{r \phi} & 0 & 0 & 0\\0 & 0 & - 2 c r^{2} & 0\\0 & 0 & 0 & - 2 c r^{2} \sin^{2}\theta\end{matrix}\right] & \left[\begin{matrix}2 \Gamma^{\phi}{}_{r \phi} & \Gamma^{t}{}_{rr} & 0 & 0\\\Gamma^{t}{}_{rr} & - 2 \Gamma^{r}{}_{rr} & 0 & 0\\0 & 0 & - 2 r \left(r \Gamma^{\phi}{}_{r \phi} - 1\right) & 0\\0 & 0 & 0 & - 2 r \left(r \Gamma^{\phi}{}_{r \phi} - 1\right) \sin^{2}{\theta}\end{matrix}\right] \end{matrix} \right. 
    \nonumber \\
    & \qquad \left. \begin{matrix} \left[\begin{matrix}0 & 0 & - \frac{c^{2} r^{2} + 1}{c} & 0\\0 & 0 & - r^{2} \Gamma^{\phi}{}_{r \phi} & 0\\- \frac{c^{2} r^{2} + 1}{c} & - r^{2} \Gamma^{\phi}{}_{r \phi} & 0 & 0\\0 & 0 & 0 & 0\end{matrix}\right] & \left[\begin{matrix}0 & 0 & 0 & - \frac{\left(c^{2} r^{2} + 1\right) \sin^{2}{\theta}}{c}\\0 & 0 & 0 & - r^{2} \Gamma^{\phi}{}_{r \phi} \sin^{2}{\theta}\\0 & 0 & 0 & 0\\- \frac{\left(c^{2} r^{2} + 1\right) \sin^{2}{\theta}}{c} & - r^{2} \Gamma^{\phi}{}_{r \phi} \sin^{2}{\theta} & 0 & 0\end{matrix}\right]\end{matrix}\right] \,,
\end{align}}\normalsize
while the nonmetricity scalar \eqref{Qscalar} is
\begin{align}
    Q &= - \frac{6 c r^{2} \Gamma^{t}{}_{rr} - 6 r^{2} \left(\Gamma^{\phi}{}_{r \phi}\right)^{2} + 6 r^{2} \Gamma^{\phi}{}_{r \phi} \Gamma^{r}{}_{rr} + 12 r \Gamma^{\phi}{}_{r \phi} - 8}{2 r^{2}} \,.
\end{align}
It is apparent that there is no choice of the free functions $\Gamma^{\phi}{}_{r \phi}$, $\Gamma^{t}{}_{rr}$, $\Gamma^{r}{}_{rr}$ that could make the nonmetricity tensor and nonmetricity scalar to vanish in the Minkowski metric case. 

\subsection{Connection set 2 branch 1}

In spherical coordinates, taking the metric \eqref{eq: spherical metric} to be Minkowski \eqref{eq: Minkowski metric}, and the connection to be given by Eq.\ \eqref{eq: Gamma set2}, the nonmetricity tensor \eqref{NonMetricityTensor} is given by
\small{\begin{align}
    Q_{\rho\mu\nu} &=\left[\begin{matrix}\left[\begin{matrix}- 2 \left(2 c^{2} \Gamma^{t}{}_{\theta \theta} - c k \Gamma^{t}{}_{\theta \theta} + c - k\right) & \frac{\left(2 c - k\right) \left(c \left(\Gamma^{r}{}_{\theta \theta}\right)^{2} + c \left(\Gamma^{t}{}_{\theta \theta}\right)^{2} + \Gamma^{t}{}_{\theta \theta}\right)}{\Gamma^{r}{}_{\theta \theta}} & 0 & 0\\\frac{\left(2 c - k\right) \left(c \left(\Gamma^{r}{}_{\theta \theta}\right)^{2} + c \left(\Gamma^{t}{}_{\theta \theta}\right)^{2} + \Gamma^{t}{}_{\theta \theta}\right)}{\Gamma^{r}{}_{\theta \theta}} & - 2 c \left(2 c \Gamma^{t}{}_{\theta \theta} - k \Gamma^{t}{}_{\theta \theta} + 1\right) & 0 & 0\\0 & 0 & - 2 c r^{2} & 0\\0 & 0 & 0 & - 2 c r^{2} \sin^{2}{\theta}\end{matrix}\right] \end{matrix} \right. \nonumber \\ 
    & \quad \left. \begin{matrix} & \left[\begin{matrix}\frac{2 \left(2 c - k\right) \left(c \Gamma^{t}{}_{\theta \theta} + 1\right) \Gamma^{t}{}_{\theta \theta}}{\Gamma^{r}{}_{\theta \theta}} & - 2 c^{2} \Gamma^{t}{}_{\theta \theta} + c k \Gamma^{t}{}_{\theta \theta} - c + \Gamma^{t}{}_{rr} & 0 & 0\\- 2 c^{2} \Gamma^{t}{}_{\theta \theta} + c k \Gamma^{t}{}_{\theta \theta} - c + \Gamma^{t}{}_{rr} & - 2 \Gamma^{r}{}_{rr} & 0 & 0\\0 & 0 & \frac{2 r \left(c r \Gamma^{t}{}_{\theta \theta} + r + \Gamma^{r}{}_{\theta \theta}\right)}{\Gamma^{r}{}_{\theta \theta}} & 0\\0 & 0 & 0 & \frac{2 r \left(c r \Gamma^{t}{}_{\theta \theta} + r + \Gamma^{r}{}_{\theta \theta}\right) \sin^{2}{\theta}}{\Gamma^{r}{}_{\theta \theta}}\end{matrix}\right] \end{matrix} \right. \nonumber \\ 
    & \quad \left. \begin{matrix} & \left[\begin{matrix}0 & 0 & - c r^{2} + \Gamma^{t}{}_{\theta \theta} & 0\\0 & 0 & \frac{c r^{2} \Gamma^{t}{}_{\theta \theta} + r^{2} - \left(\Gamma^{r}{}_{\theta \theta}\right)^{2}}{\Gamma^{r}{}_{\theta \theta}} & 0\\- c r^{2} + \Gamma^{t}{}_{\theta \theta} & \frac{c r^{2} \Gamma^{t}{}_{\theta \theta} + r^{2} - \left(\Gamma^{r}{}_{\theta \theta}\right)^{2}}{\Gamma^{r}{}_{\theta \theta}} & 0 & 0\\0 & 0 & 0 & 0\end{matrix}\right] \end{matrix} \right. \nonumber \\ 
    & \quad \left. \begin{matrix} & \left[\begin{matrix}0 & 0 & 0 & - \left(c r^{2} - \Gamma^{t}{}_{\theta \theta}\right) \sin^{2}{\theta}\\0 & 0 & 0 & \frac{\left(c r^{2} \Gamma^{t}{}_{\theta \theta} + r^{2} - \left(\Gamma^{r}{}_{\theta \theta}\right)^{2}\right) \sin^{2}{\theta}}{\Gamma^{r}{}_{\theta \theta}}\\0 & 0 & 0 & 0\\- \left(c r^{2} - \Gamma^{t}{}_{\theta \theta}\right) \sin^{2}{\theta} & \frac{\left(c r^{2} \Gamma^{t}{}_{\theta \theta} + r^{2} - \left(\Gamma^{r}{}_{\theta \theta}\right)^{2}\right) \sin^{2}{\theta}}{\Gamma^{r}{}_{\theta \theta}} & 0 & 0\end{matrix}\right]\end{matrix}\right] \,,
\label{eq: nonmetricity tensor set 2}
\end{align}}\normalsize
while the nonmetricity scalar \eqref{Qscalar} is another long expression (can be found explicitly in Refs. \cite{DAmbrosio:2021zpm,Bahamonde:2022esv}). Again, for arbitrary $c$, $k$ there is no choice of the free functions $\Gamma^{t}{}_{\theta \theta}$, $\Gamma^{t}{}_{rr}$, $\Gamma^{r}{}_{\theta \theta}$, $\Gamma^{r}{}_{rr}$ that could make the nonmetricity tensor and nonmetricity scalar to vanish in the Minkowski metric case.

\subsection{Connection set 2 branch 2}

The only way to make the nonmetricity tensor \eqref{eq: nonmetricity tensor set 2} to vanish is to set $c=k=0$ (connection set 2 branch 2) which yields
\begin{align}
    Q_{\rho\mu\nu} &=
    \left[\begin{matrix}\left[\begin{matrix}0 & 0 & 0 & 0\\0 & 0 & 0 & 0\\0 & 0 & 0 & 0\\0 & 0 & 0 & 0\end{matrix}\right] & \left[\begin{matrix}0 & \Gamma^{t}{}_{rr} & 0 & 0\\\Gamma^{t}{}_{rr} & - 2 \Gamma^{r}{}_{rr} & 0 & 0\\0 & 0 & \frac{2 r \left(r + \Gamma^{r}{}_{\theta \theta}\right)}{\Gamma^{r}{}_{\theta \theta}} & 0\\0 & 0 & 0 & \frac{2 r \left(r + \Gamma^{r}{}_{\theta \theta}\right) \sin^{2}{\theta}}{\Gamma^{r}{}_{\theta \theta}}\end{matrix}\right] \end{matrix} \right. \nonumber \\ 
    & \quad \left. \begin{matrix} & \left[\begin{matrix}0 & 0 & \Gamma^{t}{}_{\theta \theta} & 0\\0 & 0 & \frac{r^{2} - \left(\Gamma^{r}{}_{\theta \theta}\right)^{2}}{\Gamma^{r}{}_{\theta \theta}} & 0\\\Gamma^{t}{}_{\theta \theta} & \frac{r^{2} - \left(\Gamma^{r}{}_{\theta \theta}\right)^{2}}{\Gamma^{r}{}_{\theta \theta}} & 0 & 0\\0 & 0 & 0 & 0\end{matrix}\right] & \left[\begin{matrix}0 & 0 & 0 & \Gamma^{t}{}_{\theta \theta} \sin^{2}{\theta}\\0 & 0 & 0 & \frac{\left(r^{2} - \left(\Gamma^{r}{}_{\theta \theta}\right)^{2}\right) \sin^{2}{\theta}}{\Gamma^{r}{}_{\theta \theta}}\\0 & 0 & 0 & 0\\\Gamma^{t}{}_{\theta \theta} \sin^{2}{\theta} & \frac{\left(r^{2} - \left(\Gamma^{r}{}_{\theta \theta}\right)^{2}\right) \sin^{2}{\theta}}{\Gamma^{r}{}_{\theta \theta}} & 0 & 0\end{matrix}\right]\end{matrix}\right]
\end{align}
and the nonmetricity scalar
\begin{align}
    Q &= \frac{2 \left(r + \Gamma^{r}{}_{\theta \theta}\right) \left(r \Gamma^{r}{}_{\theta \theta} \Gamma^{r}{}_{rr} + r - \left(\Gamma^{r}{}_{\theta \theta}\right)^{2} \Gamma^{r}{}_{rr} + \Gamma^{r}{}_{\theta \theta}\right)}{r^{2} \left(\Gamma^{r}{}_{\theta \theta}\right)^{2}} \,.
\end{align}
Thus when in addition to the metric condition \eqref{eq: Minkowski metric} we also set
\begin{align}
\label{eq: good Gammas set 2 branch 2}
    \Gamma^{t}{}_{\theta \theta} &= 0 \,, \qquad \Gamma^{t}{}_{rr} =0 \,, \qquad  \Gamma^{r}{}_{\theta \theta} = - r  \,, \qquad \Gamma^{r}{}_{rr} = 0 \,
\end{align}
the nonmetricity tensor and scalar vanish in the Minkowski metric case. Note that the functions \eqref{eq: good Gammas set 2 branch 2} are in accord with the condition \eqref{eq: Gamma relations set22}, so everything is consistent.

What we see in this brief interlude, is that only in the connection set 2 branch 2 case the configuration is endowed with a direct Minkowski limit whereby nonmetricity can vanish. In this light perhaps it was not just a coincidence that only for set 2 branch 2 the black hole solutions were found in Refs.\ \cite{DAmbrosio:2021zpm,Bahamonde:2022esv}, while set 2 branch 1 runs into troubles with asymptotic flatness \cite{Bahamonde:2022esv}. Although further detailed investigation is welcome, it is plausible that only set 2 branch 2 connection is a  physically meaningful choice. Thus perhaps we should not be surprised that only for this branch the coincident gauge metric in the Minkowski limit turned out to be just the Minkowski metric in Cartesian coordinates. These statements, however, must be taken as conjectures and not as definite proofs stating that the other sets are not physically interesting. There are some nontrivial exact black hole solutions in other teleparallel theories where in the Minkowski limit the torsion scalar is nonvanishing but still, the configuration seems to be physically acceptable~\cite{Bahamonde:2021srr,Bahamonde:2022lvh}.

\section{Schwarzschild solution}
\label{sec:Schwarzschild}

To illustrate the general results of Sec.\ \ref{sec:coincidentG} let us consider a simple example.
When the static spherically symmetric metric \eqref{eq: spherical metric} takes the Schwarzschild form
\begin{equation}
\label{eq: Schwarzschild metric}
    g_{tt}=-\left(1-\frac{2M}{r} \right) \,, \qquad g_{rr} = \frac{1}{\left(1-\frac{2M}{r} \right)} \,,
\end{equation}
the field equations of the particular class of theories presented in~\eqref{eq: scalar-tensor field equations} are satisfied for a constant scalar field that efffectively reduces the theory to STEGR,
\begin{align}
    \mathcal{A}({\Phi}) &= 1 \,, \qquad \mathcal{B}(\Phi)= 0 \,, \qquad \mathcal{V}(\Phi) = 0 \,. \label{scalar STEGR}
\end{align}
In particular, the connection field equations are identically satisfied and any combination of the connection components is a solution, provided the curvature and torsion vanish. If we restrict to the connection that also exhibits static spherical symmetry, then we are left with to functional freedoms encoded in set 1 \eqref{eq: Gamma set1}, \eqref{eq: Gamma relations set1}, or set 2 \eqref{eq: Gamma set2}, \eqref{eq: Gamma relations set2}. Still, in the context of STEGR these two functional freedoms in the connection are left completely free by the field equations. All the same applies when we transform to the coincident gauge, as the field equations with the metric functions \eqref{eq: Schwarzschild metric} do not impose any further constraints to the  connection functions in any coordinate system. However we may attempt to fix the freedom in connection by demanding that in the coincident gauge the Schwarzschild metric should take some prescribed form. Since we assume a reasonable Minkowski limit to hold we only consider the connection set 2 branch 2 here (Eq.\ \eqref{eq: Gamma set2} with $c=k=0$).

\subsection{Cartesian form}

The first option one may naturally think of is that in the coincident gauge the 
the metric \eqref{eq: spherical metric}, \eqref{eq: Schwarzschild metric} should assume the same form as in the Cartesian coordinates, i.e. the one which is obtained by the well known definitions
\begin{subequations}
\label{eq: coinc coordinates set 22 Cartesian}
\begin{align}
\xi^0 &= t  \,,  \\
\xi^1 &= x = r \sin{\theta} \cos{\phi} \,, \\
\xi^2 &= y = r \sin{\theta} \sin{\phi} \,, \\
\xi^3 &= z = r \cos{\theta} \,,
\end{align}
\end{subequations}
in the transformation \eqref{eq: metric transformation}, yielding
\begin{align}
\label{eq: Schwarzschild in Cartesian}
    g_{\mu\nu} & = \begin{pmatrix}-(1-\frac{2 M}{r}) & 0 & 0 & 0\\0 & \frac{ r^{3}-2 M \left({y}^{2} + {z}^{2}\right)}{r^{2} \left(r-2 M\right)} & \frac{2 M {x} {y}}{r^{2} \left(r - 2 M\right)} & \frac{2 M {x} {z}}{r^{2} \left(r - 2 M\right)}\\0 & \frac{2 M {x} {y}}{r^{2} \left(r - 2 M\right)} & \frac{r^{3} - 2 M \left({x}^{2} + {z}^{2}\right)}{r^{2} \left(r - 2 M\right)} & \frac{2 M {y} {z}}{r^{2} \left(r - 2 M\right)}\\0 & \frac{2 M {x} {z}}{r^{2} \left(r - 2 M \right)} & \frac{2 M {y} {z}}{r^{2} \left(r - 2 M\right)} & \frac{r^{3} - 2 M \left({x}^{2} + {y}^{2}\right)}{r^{2} \left(r - 2 M\right)}\end{pmatrix} \,,
\end{align}
where $r^2=x^2+y^2+z^2$. The desired result is actually rather easy to obtain. The general coincident gauge metric \eqref{eq: coinc metric set 22} reduces to \eqref{eq: Schwarzschild in Cartesian} if the connection of set 2 \eqref{eq: Gamma set2} with $c=k=0$ is specified by
\begin{align}
\label{eq: good Gammas Schwarzschild Cartesian}
    \Gamma^{t}{}_{\theta \theta} &= 0 \,, \qquad \Gamma^{t}{}_{rr} =0 \,, \qquad \Gamma^{r}{}_{\theta \theta} = - r \,, \qquad \Gamma^{r}{}_{rr} = 0 \,,
\end{align}
which is exactly the same as in the Minkowski limit \eqref{eq: good Gammas set 2 branch 2}, and matches the example proposed in Refs.\ \cite{Zhao:2021zab,Lin:2021uqa}. The connection functions \eqref{eq: good Gammas Schwarzschild Cartesian} reduce the coordinate transformation \eqref{eq: coinc coordinates set 22} to \eqref{eq: coinc coordinates set 22 Cartesian}, hence obtaining the Schwarzschild metric in the Cartesian coordinates is completely anticipated. The expressions of $\Gamma^{t}{}_{rr}$ and $\Gamma^{r}{}_{rr}$ in \eqref{eq: good Gammas Schwarzschild Cartesian} follow from Eq.\ \eqref{eq: Gamma relations set22}.

Thus although the connection components are not determined by the field equations, they still seem to have certain ``natural'' values if the Cartesian form for the coincident gauge can be considered as such. For this particular set of connection components \eqref{eq: good Gammas Schwarzschild Cartesian} it turns out that the boundary term $B_Q$ \eqref{eq: B_Q} vanishes, and the GR and STEGR actions have the same value, i.e.
\begin{align}
    \lc{R} &= Q = B_Q = 0 \,.
\end{align}
While it is well known that for vacuum solutions like Scwarzschild the Levi-Civita curvature scalar $\lc{R}$ is zero, for some other completely arbitrary forms of $\Gamma^r{}_{\theta\theta}$ the nonmetricity scalar $Q$ \eqref{Qscalar} and the boundary term $B_Q$ would not be zero, although they would still compensate each other to satisfy the relation \eqref{R and Q}.

The full affine connection in symmetric teleparallelism \eqref{Connection decomposition} is composed of the Levi-Civita part \eqref{LeviCivita} and the extra part called disformation tensor \eqref{Disformation}. The latter is related to the nonmetricity tensor \eqref{NonMetricityTensor} and supplies the extra bits that make the curvature \eqref{CurvatureTensor} and torsion \eqref{TorsionTensor} of the symmetric teleparallel connection to vanish. It is instructive to compare the full symmetric teleparallel connection $\Gamma^\lambda{}_{\mu\nu}$ to the disformation $L^\lambda{}_{\mu\nu}$ in the spherical coordinates for the connection \eqref{eq: Gamma set2}, \eqref{eq: good Gammas Schwarzschild Cartesian}:
\small{\begin{align}
\label{eq: Schwarzschild Cartesian connection}
    \Gamma^\lambda{}_{\mu\nu} &= \left[\begin{matrix}\left[\begin{matrix}0 & 0 & 0 & 0\\0 & 0 & 0 & 0\\0 & 0 & 0 & 0\\0 & 0 & 0 & 0\end{matrix}\right] & \left[\begin{matrix}0 & 0 & 0 & 0\\0 & 0 & 0 & 0\\0 & 0 & - r & 0\\0 & 0 & 0 & - r \sin^{2}{\theta}\end{matrix}\right] & \left[\begin{matrix}0 & 0 & 0 & 0\\0 & 0 & \frac{1}{r} & 0\\0 & \frac{1}{r} & 0 & 0\\0 & 0 & 0 & - \sin{\theta} \cos{\theta}\end{matrix}\right] & \left[\begin{matrix}0 & 0 & 0 & 0\\0 & 0 & 0 & \frac{1}{r}\\0 & 0 & 0 & \cot{\theta}\\0 & \frac{1}{r} & \cot{\theta} & 0\end{matrix}\right]\end{matrix}\right] \,, \\
\label{eq: Schwarzschild Cartesian disformation}
    L^\lambda{}_{\mu\nu} &= \left[\begin{matrix}\left[\begin{matrix}0 & \frac{M}{r \left(2 M - r\right)} & 0 & 0\\\frac{M}{r \left(2 M - r\right)} & 0 & 0 & 0\\0 & 0 & 0 & 0\\0 & 0 & 0 & 0\end{matrix}\right] & \left[\begin{matrix}\frac{M \left(2 M - r\right)}{r^{3}} & 0 & 0 & 0\\0 & \frac{M}{r \left(r - 2 M\right)} & 0 & 0\\0 & 0 & - 2 M & 0\\0 & 0 & 0 & - 2 M \sin^{2}{\theta}\end{matrix}\right] & \end{matrix} 
    \begin{matrix} \left[\begin{matrix}0 & 0 & 0 & 0\\0 & 0 & 0 & 0\\0 & 0 & 0 & 0\\0 & 0 & 0 & 0\end{matrix}\right] & \left[\begin{matrix}0 & 0 & 0 & 0\\0 & 0 & 0 & 0\\0 & 0 & 0 & 0\\0 & 0 & 0 & 0\end{matrix}\right]\end{matrix}\right] \,.
\end{align}}\normalsize
We see that only the disformation part \eqref{eq: Schwarzschild Cartesian disformation} carries information about the Schwarzschild mass. It may be interpreted as related to the gravitational force, as in the Minkowski limit where the source of gravity disappears the disformation tensor vanishes. At the same time the full connection \eqref{eq: Schwarzschild Cartesian connection} is independent of the $M$ and has the same form also in the Minkowski case. It is tempting to interpret that as representing ``inertia'' which is present due to the curvilinear nature of the spherical coordinate system. The same features would of course carry over to the coincident gauge, i.e.\ the Cartesian coordinates, where disformation transforming as a tensor would still be proportional to $M$ (now appearing in many nonzero $x, y, z$ components), while the symmetric teleparallel connection will be zero as the curviliearity of the coordinates is ``straightened out'' removing the need to give account of the inertial effects. The Levi-Civita connection which is given by the difference of the full connection and disformation contains information about both the ``gravity'' and ``inertia'' aspect of the geometry. In GR, only the Levi-Civita connection is present which makes gravity and inertia inseparable. In STEGR these two aspects can in principle be distinguished, at least on the formal mathematical level.

Following a similar treatment as in TEGR \cite{Aldrovandi:2013wha,Krssak:2018ywd}, the last point can be illustrated by rewriting the geodesic equation
\begin{align}
    \frac{d^2 x^\rho}{ds^2} + \lc{\Gamma}^\rho{}_{\mu\nu} \frac{dx^\mu}{ds} \frac{dx^\nu}{ds} &=0
\end{align}
which desribes the motion of a unit mass particle, using the decomposition of the Levi-Civita connection into symmetric teleparallel connection and disformation \eqref{Connection decomposition},
\begin{align}
    \frac{d^2 x^\rho}{ds^2} + {\Gamma}^\rho{}_{\mu\nu} \frac{dx^\mu}{ds} \frac{dx^\nu}{ds} &= L^\rho{}_{\mu\nu} \frac{dx^\mu}{ds} \frac{dx^\nu}{ds} \,.
\end{align}
In this form it is indieed physically suggestive to interpret the LHS as describing the free particle inertial motion in curvilinear coordinates, while the RHS could be seen as representing the gravitational force, proportional to the mass of the source. In this reasoning the split of the particle equation of motion into the inertial and force parts is not arbitrary, but the the freedom in the symmetric teleparallel connection was fixed \eqref{eq: good Gammas Schwarzschild Cartesian} by the heuristic argument that the coincident gauge should correspond to the Cartesian coordinates. The main issue with this argument is that the Cartesian form is not the only one compatible with the coincident gauge, as becomes apparent below.

\subsection{Kerr-Schild form}

An alternative option also worth to entertain would be to ask whether in the coincident gauge the metric \eqref{eq: spherical metric}, \eqref{eq: Schwarzschild metric} could assume the same form as in the Kerr-Schild coordinates~\cite{Kerr:1963ud}, 
\begin{align}
\label{eq: Kerr-Schild general}
    g_{\mu \nu} &= \eta_{\mu\nu} + f \, k_\mu k_\nu \,, \qquad k_\mu = \begin{pmatrix} 1 & \frac{k_x}{\sqrt{k_x^2+k_y^2+k_z^2}} & \frac{k_y}{\sqrt{k_x^2+k_y^2+k_z^2}} & \frac{k_z}{\sqrt{k_x^2+k_y^2+k_z^2}} \end{pmatrix} \,,
\end{align}
where $f$ is a function and the vector $k_\mu$ is by construction lightlike with respect to both $\eta_{\mu\nu}$ and $g_{\mu \nu}$ \eqref{eq: Kerr-Schild general}. In the case of the Schwarzschild solution $f=\frac{2M}{r}$ and $k_i= \begin{pmatrix} \frac{x}{r} & \frac{y}{r} & \frac{z}{r} \end{pmatrix}$, and thus the metric \eqref{eq: Kerr-Schild general} written out explicitly has the form
\begin{align}
\label{eq: Schwarzschild in Kerr-Schild}
    g_{\mu\nu} &= \begin{pmatrix} - \left(1 - \frac{2 M}{r}\right) & \frac{2 M {x}}{r^{2}} & \frac{2 M {y}}{r^{2}} & \frac{2 M {z}}{r^{2}}\\\frac{2 M {x}}{r^{2}} & 1 + \frac{2 M {x}^{2}}{r^{3}} & \frac{2 M {x} {y}}{r^{3}} & \frac{2 M {x} {z}}{r^{3}}\\\frac{2 M {y}}{r^{2}} & \frac{2 M {x} {y}}{r^{3}} & 1+ \frac{2 M {y}^{2}}{r^{3}} & \frac{2 M {y} {z}}{r^{3}}\\\frac{2 M {z}}{r^{2}} & \frac{2 M {x} {z}}{r^{3}} & \frac{2 M {y} {z}}{r^{3}} & 1+ \frac{2 M {z}^{2}}{r^{3}} \end{pmatrix} \,,
\end{align}
where $r^2=x^2+y^2+z^2$. Surprisingly or not, but this result is also relatively easy to obtain. The general coincident gauge metric \eqref{eq: coinc metric set 22} reduces to \eqref{eq: Schwarzschild in Kerr-Schild} if the connection of set 2 \eqref{eq: Gamma set2} with $c=k=0$ is specified by 
\begin{align}
\label{eq: good Gammas Schwarzschild Kerr-Schild}
    \Gamma^{t}{}_{\theta \theta} &= \frac{2M}{1-\frac{2M}{r}} \,, \qquad \Gamma^{t}{}_{rr} =\frac{2M}{r^2 \left( 1-\frac{2M}{r} \right)^2} \,, \qquad \Gamma^{r}{}_{\theta \theta} = - r \,, \qquad \Gamma^{r}{}_{rr} = 0 \,.
\end{align}
Here the components $\Gamma^{t}{}_{rr}$, $\Gamma^{r}{}_{rr}$ were derived from the condition \eqref{eq: Gamma relations set22}. This connection reduces to the Minkowski limit \eqref{eq: good Gammas set 2 branch 2} when the parameter $M$ vanishes, as it should. The connection functions \eqref{eq: good Gammas Schwarzschild Kerr-Schild} give the coordinate transformation \eqref{eq: coinc coordinates set 22} the following form
\begin{subequations}
\label{eq: coinc coordinates set 22 Kerr-Schild}
\begin{align}
\xi^0 &= t + 2M \ln(r-2M) \,,  \\
\xi^1 &= x = r \sin{\theta} \cos{\phi} \,, \\
\xi^2 &= y = r \sin{\theta} \sin{\phi} \,, \\
\xi^3 &= z = r \cos{\theta} \,,
\end{align}
\end{subequations}
valid outside of the black hole horizon $r>2M$.

Although the connection components for the Cartesian form \eqref{eq: good Gammas Schwarzschild Cartesian} and Kerr-Schild form \eqref{eq: good Gammas Schwarzschild Kerr-Schild} are different, also in the latter case it turns out that the boundary term $B_Q$ \eqref{eq: B_Q} vanishes, and the GR and STEGR actions have the same value, i.e.
\begin{align}
    \lc{R} &= Q = B_Q = 0 \,.
\end{align}
At the same time the split of the Levi-Civita connection into the teleparallel connection and disformation tensor for \eqref{eq: good Gammas Schwarzschild Kerr-Schild} yields a somewhat different result from the previous case, namely
\begin{align}
\label{eq: Schwarzschild Kerr-Schild connection}
    \Gamma^\lambda{}_{\mu\nu} &= \left[\begin{matrix}\left[\begin{matrix}0 & 0 & 0 & 0\\0 & \frac{2 M}{\left(r - 2 M\right)^{2}} & 0 & 0\\0 & 0 & \frac{2 M r}{r - 2 M} & 0\\0 & 0 & 0 & \frac{2 M r \sin^{2}\theta}{r - 2 M}\end{matrix}\right] & \left[\begin{matrix}0 & 0 & 0 & 0\\0 & 0 & 0 & 0\\0 & 0 & - r & 0\\0 & 0 & 0 & - r \sin^{2}\theta\end{matrix}\right] \end{matrix} 
    \begin{matrix} \left[\begin{matrix}0 & 0 & 0 & 0\\0 & 0 & \frac{1}{r} & 0\\0 & \frac{1}{r} & 0 & 0\\0 & 0 & 0 & - \sin\theta \cos\theta\end{matrix}\right] & \left[\begin{matrix}0 & 0 & 0 & 0\\0 & 0 & 0 & \frac{1}{r}\\0 & 0 & 0 & \cot\theta\\0 & \frac{1}{r} & \cot\theta & 0\end{matrix}\right]\end{matrix}\right] \,, \\
\label{eq: Schwarzschild Kerr-Schild disformation}
    L^\lambda{}_{\mu\nu} &= \left[\begin{matrix}\left[\begin{matrix}0 & \frac{M}{r \left(2 M - r\right)} & 0 & 0\\\frac{M}{r \left(2 M - r\right)} & \frac{2 M}{\left(2 M - r\right)^{2}} & 0 & 0\\0 & 0 & \frac{2 M r}{r - 2 M} & 0\\0 & 0 & 0 & \frac{2 M r \sin^{2}\theta}{r - 2 M}\end{matrix}\right] & \left[\begin{matrix}\frac{M \left(2 M - r\right)}{r^{3}} & 0 & 0 & 0\\0 & \frac{M}{r \left(r - 2 M\right)} & 0 & 0\\0 & 0 & - 2 M & 0\\0 & 0 & 0 & - 2 M \sin^{2}\theta\end{matrix}\right] \end{matrix} 
    \begin{matrix} \left[\begin{matrix}0 & 0 & 0 & 0\\0 & 0 & 0 & 0\\0 & 0 & 0 & 0\\0 & 0 & 0 & 0\end{matrix}\right] & \left[\begin{matrix}0 & 0 & 0 & 0\\0 & 0 & 0 & 0\\0 & 0 & 0 & 0\\0 & 0 & 0 & 0\end{matrix}\right]\end{matrix}\right] \,.
\end{align}
It is not immediately clear what conclusion can be drawn from here. While the disformation tensor is still proportional to the mass $M$, there are also components containing $M$ in the teleparallel connection. On the one hand one may think that this feature is unnatural, and the Cartesian form is more true to physical intuition. On the other hand the metric contains the parameter $M$ as well, and it is perhaps a bit premature to claim that $M$ in the connection \eqref{eq: Schwarzschild Kerr-Schild disformation} can not measure any inertial effects, especially as they all get eliminated in the coincident gauge, i.e.\ in the Kerr-Schild coordinates where the symmetric teleparallel connection vanishes. It goes without saying that in the Minkowski limit \eqref{eq: Minkowski metric} both connections \eqref{eq: good Gammas Schwarzschild Cartesian} and \eqref{eq: good Gammas Schwarzschild Kerr-Schild} reduce to the same \eqref{eq: good Gammas set 2 branch 2}.

\subsection{Diagonal form}

After seeing the Cartesian and Kerr-Schild forms of the coincident gauge metric, an obvious question arises whether a diagonal form of the coincident gauge metric is also possible. That would require taking 
\begin{align}
\label{eq: good Gammas Schwarzschild diagonal}
    \Gamma^{t}{}_{\theta \theta} &= 0 \,, \qquad \Gamma^{t}{}_{rr} =0 \,, \qquad \Gamma^{r}{}_{\theta \theta} = -\sqrt{r (r-2M)} \,, \qquad \Gamma^{r}{}_{rr} = -\frac{(r-M)-\sqrt{r(r-2M)}}{r(r-2M)} \,
\end{align}
to make the off-diagonal components of \eqref{eq: coinc metric set 22} to vanish and satisfy the condition \eqref{eq: Gamma relations set22}. The integral in \eqref{eq: coinc coordinates set 22} that relates the radial coordinate $r$ to the coincident spatial coordinates,
\begin{align}
\label{eq: Schwarzschild R isotropic}
    R &= \sqrt{\xi_x^2+\xi_y^2+\xi_z^2} = e^{-\int \frac{1}{\Gamma^r_{\theta\theta}} dr} = e^{2 \arccosh \sqrt{\frac{r}{2M}}} =  \frac{2M}{(\sqrt{r}+\sqrt{r-2M})^{2} } \,
\end{align}
is valid only for $\frac{r}{2M} >1$, i.e.\ outside of the horizon (at $R=1$). With this restriction the coordinate transformation \eqref{eq: coinc coordinates set 22} acquires the following form
\begin{subequations}
\label{eq: coinc coordinates set 22 diagonal}
\begin{align}
\xi^0 &= t \,,  \\
\xi^1 &= \xi_x = R \, \sin{\theta} \cos{\phi} \,, \\
\xi^2 &= \xi_y = R \, \sin{\theta} \sin{\phi} \,, \\
\xi^3 &= \xi_z = R \, \cos{\theta} \,,
\end{align}
\end{subequations}
and the metric turns out to be 
\begin{align}
\label{eq: Schwarzschild in diagonal}
      g_{\mu\nu} &=\left(
\begin{array}{cccc}
 -\Big(\frac{R-1}{R+1}\Big)^2 & 0 & 0 & 0 \\
 0 & \frac{1}{4} M^2 \left(1+\frac{1}{R}\right)^4 & 0 & 0 \\
 0 & 0 & \frac{1}{4} M^2 \left(1+\frac{1}{R}\right)^4 & 0 \\
 0 & 0 & 0 & \frac{1}{4} M^2 \left(1+\frac{1}{R}\right)^4 \\
\end{array}
\right)\,. 
\end{align}
The structure of this form reminds the Schawarzchild metric in isotropic coordinates, however the multiplicative position of the parameter $M$ is different. Despite this, when the relation between $R$ and $M$ \eqref{eq: Schwarzschild R isotropic} is properly taken into account the metric \eqref{eq: Schwarzschild in diagonal} reduces to the Cartesian form $\eta_{\mu\nu}$ in the Minkowski limit $M \to 0$, as expected.

In contrast to the two previous cases the nonmetricity scalar $Q$ given by Eq.~\eqref{Qscalar} (and correspondingly also the boundary term $B_Q$ given by~\eqref{eq: B_Q}) computed in spherical coordinates for the metric \eqref{eq: Schwarzschild metric} and connection \eqref{eq: good Gammas Schwarzschild diagonal} is nontrivial, namely,
\begin{align}
    Q &= - B_Q = \frac{4 \left(- 2 M^{2} r^{\frac{5}{2}} + 3 M r^{\frac{7}{2}} - 2 M r^{3} \sqrt{r - 2 M} - r^{\frac{9}{2}} + r^{4} \sqrt{r - 2 M}\right)}{r^{5} \left(r - 2 M\right)^{\frac{3}{2}}}\,.
\end{align}
In other words, the choice of the connection \eqref{eq: good Gammas Schwarzschild diagonal} means the GR action of $\lc{R}$ and the STEGR action of $Q$ differ numerically (by the boundary term), although their differential structure is equivalent in the sense that its variation gives the same field equations.

Just for reference and comparison, let us also give here the teleparallel connection and disformation tensor that arise from the connection \eqref{eq: Schwarzschild R isotropic} in spherical coordinates:
\small{\begin{align}
\label{eq: Schwarzschild isotropic connection}
    \Gamma^\lambda{}_{\mu\nu} &= \left[\begin{matrix} \left[\begin{matrix}0 & 0 & 0 & 0\\0 & 0 & 0 & 0\\0 & 0 & 0 & 0\\0 & 0 & 0 & 0\end{matrix}\right] & \left[\begin{matrix}0 & 0 & 0 & 0\\0 & - \frac{r - M - \sqrt{r (r - 2 M)}}{r \left(r - 2 M\right)} & 0 & 0\\0 & 0 & - \sqrt{r(r - 2 M)} & 0\\0 & 0 & 0 & - \sqrt{r ( r - 2 M ) } \sin^{2}\theta\end{matrix}\right] \end{matrix} \right. \nonumber \\
    & \qquad \qquad \left. \begin{matrix} \left[\begin{matrix}0 & 0 & 0 & 0\\0 & 0 & \frac{1}{\sqrt{r ( r - 2 M ) }} & 0\\0 & \frac{1}{\sqrt{r ( r - 2 M ) }} & 0 & 0\\0 & 0 & 0 & - \sin\theta \cos\theta\end{matrix}\right]
    & \left[\begin{matrix}0 & 0 & 0 & 0\\0 & 0 & 0 & \frac{1}{\sqrt{r ( r - 2 M ) }}\\0 & 0 & 0 & \cot\theta\\0 & \frac{1}{\sqrt{r ( r - 2 M ) }} & \cot\theta & 0\end{matrix}\right] \end{matrix}\right] \,, \\
\label{eq: Schwarzschild isotropic disformation}
    L^\lambda{}_{\mu\nu} &= \left[\begin{matrix} \left[\begin{matrix}0 & - \frac{M}{r \left(r - 2 M\right)} & 0 & 0\\- \frac{M}{r \left(r - 2 M\right)} & 0 & 0 & 0\\0 & 0 & 0 & 0\\0 & 0 & 0 & 0\end{matrix}\right] & \end{matrix} 
    \begin{matrix}\left[\begin{matrix}- \frac{M \left(r - 2 M\right)}{r^{3}} & 0 & 0 & 0\\0 & - \frac{r - 2 M - \sqrt{r ( r - 2 M ) }}{r \left(r - 2 M\right)} & 0 & 0\\0 & 0 & r - 2 M - \sqrt{r ( r - 2 M ) } & 0\\0 & 0 & 0 & \left(r - 2 M - \sqrt{r ( r - 2 M ) }\right) \sin^{2}\theta\end{matrix}\right] \end{matrix} \right. \nonumber \\ 
    & \qquad \qquad \left. \begin{matrix} \left[\begin{matrix}0 & 0 & 0 & 0\\0 & 0 & \frac{\sqrt{r}- \sqrt{r ( r - 2 M ) } }{r \sqrt{r - 2 M}} & 0\\0 & \frac{\sqrt{r}- \sqrt{r ( r - 2 M ) }}{r \sqrt{r - 2 M}} & 0 & 0\\0 & 0 & 0 & 0\end{matrix}\right] & \left[\begin{matrix}0 & 0 & 0 & 0\\0 & 0 & 0 & \frac{\sqrt{r}- \sqrt{r ( r - 2 M ) }}{r \sqrt{r - 2 M}}\\0 & 0 & 0 & 0 \\0 & \frac{\sqrt{r}- \sqrt{r ( r - 2 M ) }}{r \sqrt{r - 2 M}} & 0 & 0\end{matrix}\right] \end{matrix}\right] \,.
\end{align}}\normalsize
The result looks more messy than in the two previous cases, as can be expected from the more elaborate form of the connection functions \eqref{eq: good Gammas Schwarzschild diagonal}. There is no clear separation between the inertial connection without $M$ and gravity represented by the disformation proportional to $M$. However, like in the Kerr-Schild case, it may be argued that since the parameter $M$ appears in the metric \eqref{eq: Schwarzschild metric}, the inertial effects should depend on $M$ too. In the Minkowski limit $M \to 0$ the disformation tensor \eqref{eq: Schwarzschild Kerr-Schild disformation} vanishes, while the teleparallel connection \eqref{eq: Schwarzschild isotropic connection} reduces to the same form as in the Cartesian case \eqref{eq: Schwarzschild Cartesian connection}.

In summary, we saw how the coincident gauge for the Schwarzschild solution was not unique, and different choices of the coincident gauge coordinates are possible. Prescribing certain form for the coincident gauge metric fixed the functional freedom in the teleparallel connection, that was otherwise left undetermined by the field equations. Yet, without some extra input or argument is is hard to decide which form of the teleparallel connection could be considered truly physical (either \eqref{eq: good Gammas Schwarzschild Cartesian}, \eqref{eq: good Gammas Schwarzschild Kerr-Schild}, \eqref{eq: good Gammas Schwarzschild diagonal}, or some other). As different choices for the connection lead to different forms of the boundary term $B_Q$, one should probably consider quantities depending on the boundary term to properly address this issue. Such investigation, however, remains beyond the scope of the present paper.

Finally let us also note that while in the coincident gauge the teleparallel connection vanishes, the resulting metrics \eqref{eq: Schwarzschild in Cartesian}, \eqref{eq: Schwarzschild in Kerr-Schild}, and \eqref{eq: Schwarzschild in diagonal} have many nontrivial components or at least a nontrivial coordinate dependence, and thus do not lead to easier computations in practice. The spherical coordinates which are well adapted to the symmetry of the configuration are more economical for practical purposes, despite being associated with a few nonzero connection components.

\section{BBMB solution}\label{sec:BBMB}

As shown in \cite{Bahamonde:2022esv} the equations \eqref{eq: scalar-tensor field equations} for the theory
\begin{align}
    \mathcal{A}({\Phi}) &= -\frac{\beta {\Phi}^2}{8}\,, \qquad \mathcal{B}(\Phi)= \beta \,, \qquad \mathcal{V}(\Phi) = 0 \, \label{scalar1simple}
\end{align}
and in the case of set 2 branch 2 connection are solved by a Bocharova-Bronnikov-Melnikov--Bekenstein (BBMB) asymptotically flat black hole metric \cite{bocharova1970exact,Bekenstein:1974sf,Bekenstein:1975ts} along with the scalar field profile
\begin{align}
\label{eq: BBMB metric spherical coordinates}
    g_{tt} &= \left(1- \frac{M}{r} \right)^2 \,, \qquad g_{rr} = \frac{1}{\left(1- \frac{M}{r}\right)^2} \,, \qquad {\Phi}(r)=\Phi_0\Big(1-\frac{M}{r}\Big)^{-1/2} \,,
\end{align}
and connection components
\begin{align}
\label{eq: Gamma_rthetatheta BBMB}
    \Gamma^r{}_{\theta\theta} = -(r-M) \,, \qquad \Gamma^r{}_{rr} = 0 \,
\end{align}
(the latter component is fixed by the condition \eqref{eq: Gamma relations set2}). In contrast to the Schwarszschild solution in STEGR discussed in the previous section, only the components $\Gamma^t{}_{\theta\theta}$ and $\Gamma^t{}_{rr}$ are left undetermined in the field equations. The coincident gauge metric \eqref{eq: coinc metric set 22} thus has one functional freedom $\Gamma^t{}_{\theta\theta}$, and fixing it in any way gives a valid metric in the coordinate system where the teleparallel connection vanishes everywhere in spacetime. In the next section we will give a simple example.

\subsection{Diagonal form}

A fairly easy choice would be to ask that the time coordinate does not get mixed with the rest, i.e.\ $\xi_w=t$, and accordingly 
\begin{align}
\label{eq: Gamma_trr BBMB}
    \Gamma^t{}_{\theta\theta} & = 0 \,, \qquad \Gamma^t{}_{rr} =0 \,.
\end{align}
This simplifies the transformation \eqref{eq: coinc coordinates set 22} a lot, giving
\begin{subequations}
\label{eq: coinc coordinates set 22 BBMB}
\begin{align}
\xi^0 &= t  \,,  \\
\xi^1 &= \xi_x = R \sin{\theta} \cos{\phi} \,, \\
\xi^2 &= \xi_y = R \sin{\theta} \sin{\phi} \,, \\
\xi^3 &= \xi_z = R \cos{\theta} \,,
\end{align}
\end{subequations}
where $R=r-M$ measures the radial distance from the black hole horizon. After transforming to the coincident coordinates \eqref{eq: coinc coordinates set 22 BBMB} the metric \eqref{eq: BBMB metric spherical coordinates} takes a bit surprisingly a diagonal form 
\begin{align}
\label{eq: BBMB metric in coincident}
    g_{\mu\nu} & = \begin{pmatrix}-\frac{1}{\left(1+\frac{M}{R}\right)^2} & 0 & 0 & 0\\0 & \left(1+\frac{M}{R}\right)^2 & 0 & 0 \\0 & 0 & \left(1+\frac{M}{R}\right)^2 & 0 \\0 & 0 & 0 & \left(1+\frac{M}{R}\right)^2 \end{pmatrix} \,,
\end{align}
which reminds of isotropic coordinates. 

For the record, let us note that the Levi-Civita Ricci scalar, nonmetricity scalar \eqref{Qscalar}, and the boundary term \eqref{eq: B_Q} are
\begin{align}
    \lc{R} &= 0 \,, \qquad Q=-\frac{2M^2}{r^4} \,, \qquad B_Q=\frac{2M^2}{r^4} \,.
\end{align}
The BBMB metric was originally found as a solution in the usual metric and torsion free scalar-tensor theory with a vanishing potential but a different form of the coupling function  $\mathcal{A}(\Phi)$ to the curvature scalar $\lc{R}$. Deeper reasons why this solution exists in the scalar-tensor versions of symmetric teleparallelism based on nonmetricity \cite{Bahamonde:2022esv} as well as metric teleparallelism based on torsion \cite{Bahamonde:2022lvh} remain to be understood.

We may also substitute the functions \eqref{eq: Gamma_rthetatheta BBMB} and \eqref{eq: Gamma_trr BBMB} into the full connection \eqref{eq: Gamma set2} and compare it with the disformation,
\small{\begin{align}
\label{eq: BBMB isotropic connection}
    \Gamma^\lambda{}_{\mu\nu} &= \left[\begin{matrix}\left[\begin{matrix}0 & 0 & 0 & 0\\0 & 0 & 0 & 0\\0 & 0 & 0 & 0\\0 & 0 & 0 & 0\end{matrix}\right] & \left[\begin{matrix}0 & 0 & 0 & 0\\0 & 0 & 0 & 0\\0 & 0 & - R & 0\\0 & 0 & 0 & - R \sin^{2}{\theta}\end{matrix}\right] & \left[\begin{matrix}0 & 0 & 0 & 0\\0 & 0 & \frac{1}{R} & 0\\0 & \frac{1}{R} & 0 & 0\\0 & 0 & 0 & - \sin{\theta} \cos{\theta}\end{matrix}\right] & \left[\begin{matrix}0 & 0 & 0 & 0\\0 & 0 & 0 & \frac{1}{R}\\0 & 0 & 0 & \cot{\theta}\\0 & \frac{1}{R} & \cot{\theta} & 0\end{matrix}\right]\end{matrix}\right] \,, 
    \\
\label{eq: BBMB isotropic disformation}
    L^\lambda{}_{\mu\nu} &= \left[\begin{matrix}\left[\begin{matrix}0 & - \frac{M}{R \left(M + R\right)} & 0 & 0\\- \frac{M}{R \left(M + R\right)} & 0 & 0 & 0\\0 & 0 & 0 & 0\\0 & 0 & 0 & 0\end{matrix}\right] & \left[\begin{matrix}- \frac{M R^{3}}{\left(M + R\right)^{5}} & 0 & 0 & 0\\0 & \frac{M}{R \left(M + R\right)} & 0 & 0\\0 & 0 & - \frac{M R}{M + R} & 0\\0 & 0 & 0 & - \frac{M R \sin^{2}{\theta}}{M + R}\end{matrix}\right] & \end{matrix} 
    \begin{matrix} \left[\begin{matrix}0 & 0 & 0 & 0\\0 & 0 & \frac{M}{R \left(M + R\right)} & 0\\0 & \frac{M}{R \left(M + R\right)} & 0 & 0\\0 & 0 & 0 & 0\end{matrix}\right] & \left[\begin{matrix}0 & 0 & 0 & 0\\0 & 0 & 0 & \frac{M}{R \left(M + R\right)}\\0 & 0 & 0 & 0\\0 & \frac{M}{R \left(M + R\right)} & 0 & 0\end{matrix}\right]\end{matrix}\right]\,.
\end{align}}\normalsize
Here the teleparallel connection can be understood as representing ``inertia'', while the disformation part of the connection represents ``gravity'', since taking $M$ to zero makes it to vanish completely. Note that the connection \eqref{eq: BBMB isotropic connection} is in an analogous form to the Schwarzschild connection in the Cartesian case \eqref{eq: Schwarzschild Cartesian connection}, but the respective expressions \eqref{eq: Schwarzschild Cartesian disformation} and \eqref{eq: BBMB isotropic disformation} for the disformation tensor have some differences. Otherwise, the coordinate transformation \eqref{eq: coinc coordinates set 22 BBMB} and the diagonal metric \eqref{eq: BBMB metric in coincident} are more similar to the transformation \eqref{eq: coinc coordinates set 22 diagonal} and the diagonal metric \eqref{eq: Schwarzschild in diagonal} of the Schwarzschild solution, but the BBMB connection \eqref{eq: Gamma_rthetatheta BBMB}, \eqref{eq: Gamma_trr BBMB} is rather different from the corresponding Schwarzschild one \eqref{eq: Schwarzschild isotropic connection}.

\subsection{Cartesian and Kerr-Schild forms are not compatible with the coincident gauge}

While in the case of the Schwarzschild solution the static spherically symmetric connection was endowed with two functional freedoms which could be used to give the coincident gauge metric a specific form, in the BBMB case one of those functional freedoms is fixed by the connection field equation and thus the options to give the coincident gauge metric an arbitrary form are more limited. The reason for this is the fact that the BBMB solution exists in the symmetric teleparallel scalar-tensor  gravity theory (see the action~\eqref{Action}) and then, the role of the extra components of the connection play a dynamical role in the field equations. Due to this, some connection components are directly set by the solving the field equations~\eqref{eq: scalar-tensor field equations}-\eqref{ScalarFieldEq}, while in the Schwarzschild case, the theory is just reduced to STEGR and the extra components of the connection do not enter into the field equations (so that they are free functions). 

One conclusion that can be obtained is that it turns out the Cartesian and Kerr-Schild forms of the metric are not consistent with the coincident gauge in the BBMB case. 
It is rather straightforward to check this claim for the Cartesian form. One would take the metric \eqref{eq: coinc metric set 22} with $\Gamma^{r}{}_{\theta \theta}$ expressed by \eqref{eq: Gamma_rthetatheta BBMB}. In order to make the components $g_{it}$ to vanish, it is necessary to require that $\Gamma^{t}{}_{\theta \theta} =0$. But this immediately brings the metric into the isotropic-like form \eqref{eq: BBMB metric in coincident}, not the Cartesian form.
To check the incompatibility of the Kerr-Schild form, one can again begin with the coincident gauge metric \eqref{eq: coinc metric set 22} with $\Gamma^{r}{}_{\theta \theta}$ given by \eqref{eq: Gamma_rthetatheta BBMB}. Equating the components of this metric with the general Kerr-Schild form \eqref{eq: Kerr-Schild general} yields an overdetermined system of equations which has no real solutions for the remaining functions $f$, $k_i$, and $\Gamma^{t}{}_{\theta \theta}$. For example, the $0i$-components give $\Gamma^t{}_{\theta\theta} =\frac{2 M r^2}{\left(r-M\right)^2}$ which in turn is incompatible with the $ij$-components. Thus for the BBMB solution of the symmetric teleparallel scalar-tensor theory considered in the coincident gauge metric can not be put into the Kerr-Schild form either. 
Of course, it is possible to transform the BBMB solution into the Cartesian or Kerr-Schild coordinates, just the teleparallel connection is not zero in those coordinates.


\section{Conclusions and discussion}\label{sec:conclusions}
In this paper we studied how the two different sets describing static spherically symmetric spacetime configurations in symmetric teleparallel gravities appear in the coincident gauge. By doing this, instead of having a metric and a nonzero affine connection, one chooses a specific coordinate system (coincident gauge) such that the connection vanishes but all the extra information coming from the connection is encoded in the metric. We have found that this method does not give a unique coordinate system, instead, there are infinite coordinate systems satisfying this condition. We solved the transformation for those coordinate systems in a generic way and then provided different particular coordinate systems as examples. In principle, this can be considered as an algorithm that one can use to derive coincident gauge coordinates. It applies not only for STEGR, but also in any modified version of symmetric teleparallel gravity as well.

As a byproduct of the investigations, we noticed that only the connection of set 2 branch 2 can provide a vanishing nonmetricity tensor and nonmetricity scalar in the Minkowski limit. For all the other branches, it is not possible to find any suitable value for the connection components such that these quantites vanish. To unpack the context, recall that in the Riemannian case, if the metric is the Minkowski one, all curvature quantities are trivially zero. However, in symmetric teleparallel gravity, since the metric and the connection are independent (or not directly related), one can have the situation that even if the metric is Minkowski, still, the nonmetricity tensor (and its scalars) could be nonvanishing. This is related to the fact that we only imposed that the connection and the metric are spherically symmetric, but it does not imply that in the Minkowski limit both quantities would still respect the same symmetries (Minkowski symmetries). It is kind of expected to have the property that all gravitational scalars should vanish in Minkowski, and one might think that due to this, one might need to discard the other branches as unphysical. However, since matter is only coupled to the metric (and at most to the Levi-Civita connection), \textit{a priori}, there is nothing unphysical to consider that even in Minkowski, those scalars (and tensors) are nonvanishing. Further, several solutions have been recently found in torsional teleparallel gravity whose scalars are nonvanishing in the Minkowski limit, and still those solutions represent physically realistic black hole configurations~\cite{Bahamonde:2022lvh,Bahamonde:2021srr,Jusufi:2022loj}. In some ways the situation is analogous to how only one branch of the axially symmetric torsional connections has a proper spherically symmetric limit \cite{Bahamonde:2020snl}.

We also studied the case of the Schwarzschild and the BBMB metrics for several coordinate systems compatible with the coincident gauge. By fixing the freedom in the connection allowed by the symmetry and the field equations, the Schwarzshild metric in the coincident gauge can take for example the Cartesian, Kerr-Schild and diagonal (isotropic-like) forms, while the BBMB metric can be put only in the diagonal but not in the Cartesian or Kerr-Schild forms. In principle, different connections imply different values for the boundary term, but there is no known well defined way to fix the undetermined connection components in those solutions. Going to the coincident gauge does not seem to provide a clear way to fix them either. We considered several arguments about trying to fix those components, ranging from the form of the coincident gauge metric to asking the teleparallel connection to be purely ``inertial'', but it feels some extra input or argument is necessary. Thus, further studies about different physical implications of those components need to be conducted. A possible way ahead could be related to choosing the so-called canonical frame of symmetric teleparallel gravity~\cite{BeltranJimenez:2019bnx,Gomes:2022vrc}, defined as an inertial frame. 

We also noticed that even though the connection vanishes in the coincident gauge, in spherical symmetry, computations do not become simpler since the metric becomes more complicated. However, it could be that one might still find some particular coordinate systems compatible with the coincident gauge where the metric becomes or remains simple. We have noticed that this can happen for certain coordinate system such as the one used in the BBMB section (see Sec.~\ref{sec:BBMB}) which resembles some notions from isotropic coordinates. Remarkably, this coordinate choice makes the metric  diagonal and very simple while the connection is vanishing. Even though the solution and its physical interpretation will not change by making those coordinate transformations, the role of the connection might be different in each of them and it could be important for studies concerning the entropy of black holes. Further investigations regarding those topics could be considered in the future.

\subsection*{Acknowledgements}S.B. is supported by JSPS Postdoctoral Fellowships for Research in Japan and KAKENHI Grant-in-Aid for Scientific Research No. JP21F21789. L.J.\ is supported by the Estonian Research Council grant PRG356 ``Gauge Gravity". The authors also acknowledge support by the European Regional Development Fund through the Center of Excellence TK133 ``The Dark Side of the Universe".

\bibliographystyle{utphys.bst}    
\bibliography{Coincident_gauge_paper_epjc}

\end{document}